\documentclass{aa}
\usepackage[utf8]{inputenc}

\usepackage{graphicx,ulem}
\usepackage{longtable}
\usepackage{booktabs}
\usepackage{float}
\usepackage{dcolumn}
\usepackage{graphics,epsfig}
\usepackage{amsmath,amssymb,latexsym,mathrsfs}
\usepackage[breaklinks, colorlinks, citecolor=blue]{hyperref}
\usepackage{bm}
\usepackage{color}
\usepackage{subfigure}
\usepackage{multirow}
\usepackage{soul}

\newcommand\morehorsp{\rule[-3mm]{0mm}{8mm}}

\usepackage{tikz}

\title{Impact of systematics on cosmological parameters from future Galaxy Clusters surveys}
\author{Laura Salvati\inst{1}\inst{2}\inst{3}\inst{*} \and Marian Douspis\inst{1} \and Nabila Aghanim\inst{1}}
\institute{\inst{1} Université Paris-Saclay, CNRS,  Institut d'Astrophysique Spatiale, 91405, Orsay, France \\
\inst{2} INAF - Osservatorio Astronomico di Trieste, via G. B. Tiepolo 11, I-34143 Trieste, Italy \\
\inst{3} IFPU - Institute for Fundamental Physics of the Universe, Via Beirut 2, 34014 Trieste, Italy\\
\inst{*}\email{laura.salvati@inaf.it}}
\authorrunning{Salvati et al.}
\date{}

\abstract{
Galaxy clusters are a recent cosmological probe. The precision and accuracy of the cosmological parameters inferred from these objects are affected by the knowledge of cluster physics, entering the analysis through the mass-observable scaling relations, and the theoretical description of their mass and redshift distribution, modelled by the mass function. In this work, we forecast the impact of different modelling of these ingredients for clusters detected by future optical and near-IR surveys. We consider the standard cosmological scenario and the case with a time-dependent equation of state for dark energy.
We analyse the effect of increasing accuracy on the scaling relation calibration, finding improved constraints on the cosmological parameters. This higher accuracy exposes the impact of the mass function evaluation, which is a subdominant source of systematics for current data. We compare two different evaluations for the mass function. In both cosmological scenarios, the use of different mass functions leads to biases in the parameter constraints. For the $\Lambda$CDM model, we find a $1.6 \, \sigma$ shift in the $(\Omega_m,\sigma_8)$ parameter plane and a discrepancy of $\sim 7 \, \sigma$ for the redshift evolution of the scatter of the scaling relations. For the scenario with a time-evolving dark energy equation of state, 
the assumption of different mass functions 
results in a $\sim 8 \, \sigma$ tension in the $w_0$ parameter. These results show the impact, and the necessity for a precise modelling, of the interplay between the redshift evolution of the mass function and of the scaling relations in the cosmological analysis of galaxy clusters.
}

\begin{document}

\maketitle

\section{Introduction} 

In the cosmological hierarchical scenario, galaxy clusters form in the recent Universe, from the collapse of high density fluctuations. The formation and evolution of these objects is strictly related to the growth history of the large scale structure and to the underlying cosmological model. For this reason, in recent years galaxy clusters have emerged as a powerful cosmological probe. 

Different wavelength observations provide catalogs of hundreds of objects to be used for the cosmological analysis, such as 
\cite{Ade:2015fva}, \cite{deHaan:2016qvy}, \cite{Bocquet:2018ukq} in the mm wavelengths, \cite{Abbott:2020knk} in optical, \cite{Boehringer:2017wvr}, \cite{Pacaud:2018zsh} in X-rays.
These different analysis show that the accuracy and precision on the cosmological parameter constraints are affected by systematic uncertainties related to the modelling of different theoretical and observational ingredients. 

In general, galaxy cluster number counts are used as a cosmological probe.
In the ideal scenario, the number counts should coincide with the halo mass function, i.e. the number distribution of clusters in bins of redshift and mass. 
However, cluster masses cannot be measured directly. It is therefore necessary to rely on observables which act as mass-proxies and that are tightly correlating with the underlying cluster mass, via some statistical scaling relation. The calibration of these scaling relations (the so-called mass-calibration problem), represents the current limiting systematic in cluster cosmology studies. 
Scaling relations are then used, together with a model of the selection process, to transform the theoretical mass function into a prediction for the distribution of clusters in the space of survey observables.

In this scenario, the halo mass function itself may be a further source of systematics.
The calibration of the mass function is usually obtained through numerical simulations. 
In the last decades many authors provided different formulations and calibrations that can be used in the cosmological analysis, see e.g. the discussion in \cite{Monaco:2016pys} and reference therein.
Nevertheless, it has been shown that the mass function calibration can impact the final results on cosmological parameters up to $\sim 10\%$, see e.g. discussion in \cite{PhysRevD.90.023520}, \cite{Bocquet:2015pva}, \cite{Bocquet:2020tes}. Indeed, the fitting formulas obtained from numerical simulations may change when considering different analysis, depending on the initial conditions and assumptions performed during the simulations (e.g. assumed initial cosmology, definition of the cluster mass and detection, resolution of the simulation, etc ..).

From the cosmological analysis of recently observed cluster samples, the scaling relation calibration stands out as the major source of systematics. 
However, in the near future different surveys will provide samples of thousands of well characterised clusters. This large statistics and the availability of multi-wavelength observations will likely improve the precision and accuracy in the calibrations of the scaling relations, reducing the impact on the cosmological parameters. It is time therefore to focus on the other ingredients entering in the analysis, e.g. the mass function. 

In this paper, we study the impact of these different systematic sources on the cosmological parameters constrained from galaxy clusters. In particular, we analyse the effect of increasing accuracy for the scaling relation calibrations and different formulations for the mass function. 
We build the entire cosmological pipeline and simulate observations from future surveys, providing results for a Euclid-like, a LSST-like and a WFIRST-like experiment.
In this way, we are also able to quantify the impact of different observation strategies, such as the observed area, the covered redshift range, etc.  

The paper is structured as follows. In section \ref{sec:method} we describe the method we adopt in the analysis, presenting our results in section \ref{sec:results}. We discuss our findings and draw the conclusions in sections \ref{sec:discussion} and \ref{sec:conclusions}.

\section{Method}\label{sec:method}

In this work we study the impact on cosmological constraints inferred from galaxy clusters of systematic effects arising from the uncertainty of the scaling relation calibrations and the choice of halo mass function in the analysis. We consider the galaxy cluster number counts as our observable. We define clusters within the radius $R_{200}$, such that the cluster mean mass over-density is 200 times the critical density at that redshift, $\rho_c(z)$. It implies that the cluster mass is defined as 
\begin{equation}\label{eq:mass}
M_{200} = \frac{4}{3} \pi R^3_{200} \rho_c(z) \, .
\end{equation}

In this section we describe the theoretical model adopted to evaluate the cluster number counts and the experimental characteristics used to simulate mock data. We then describe the fitting procedure, through a Monte Carlo Markov Chains analysis.

\subsection{Galaxy Cluster Number Counts}
For the evaluation of galaxy cluster number counts, we follow the analysis in \cite{Sartoris:2015aga}. The expected cluster number counts in a given redshift and observed mass ($M^{\text{ob}}_{200}$) bin, $ N_{\ell,m}$, for a survey with a sky coverage $\Omega _{\text{sky}}$, is defined as

\begin{eqnarray}\label{eq:s1}
N_{\ell,m} &=& \Delta \Omega _{\text{sky}} \int _{z_{\ell}}^{z_{\ell+1}}  dz \dfrac{dV}{dz d\Omega} \int _{M_{\ell,m}^{\text{ob}}}^{M_{\ell,m+1}^{\text{ob}}} \dfrac{dM^{\text{ob}}_{200}}{M^{\text{ob}}_{200}} \notag \\
&\times &  \int _0 ^{+\infty} dM \, \dfrac{dn(M_{200},z)}{dM_{200}} \, p(M^{\text{ob}}_{200}|M_{200}) \, .
\end{eqnarray}

\noindent In Eq.~\ref{eq:s1} $dV/dz d\Omega$ is the comoving volume element per unit of redshift and solid angle, $dn(M_{200},z)/dM_{200}$ is the halo mass function and $p(M^{\text{ob}}_{200}|M_{200})$ is the probability of a galaxy cluster with true mass $M$ to have an observed mass $M^{\text{ob}}$. We follow \cite{Lima:2005tt} and assume a log-normal probability density, such that
\begin{equation}\label{eq:s2}
p(M^{\text{ob}}_{200}|M_{200}) = \dfrac{\exp{\left[ -x^2(M^{\text{ob}}_{200}) \right]}}{\sqrt{2 \pi \sigma ^2 _{\ln M_{200}}}} \, ,
\end{equation}
with 
\begin{equation}\label{eq:s3}
x(M^{\text{ob}}_{200}) = \dfrac{\ln M^{\text{ob}}_{200} - \ln M_{\text{bias}} - \ln M_{200}}{\sqrt{2 \sigma ^2 _{\ln M_{200}}}} \, .
\end{equation}

\noindent The combination of Eqs.~\ref{eq:s1} and \ref{eq:s2} provides
\begin{eqnarray}\label{eq:s4}
N_{\ell,m} &=& \dfrac{\Delta \Omega _{\text{sky}}}{2} \int _{z_{\ell}} ^{z_{\ell +1}} dz \dfrac{dV}{dz d\Omega}  \notag \\
& \times & \int _{0} ^{+\infty} dM_{200} \,  \dfrac{dn(M_{200},z)}{dM_{200}} \notag  \\
& \times & \left[ \text{erfc}(x(M^{\text{ob}}_{\ell, m})) - \text{erfc}(x(M^{\text{ob}}_{\ell, m+1})) \right] \, ,
\end{eqnarray}
with $\text{erfc}(x)$ being the complementary error function.

The definition of $x(M^{\text{ob}}_{200})$ provides the link with the scaling relations, with $\ln M_{\text{bias}}$ being the bias in the mass estimation
\begin{equation}\label{eq:s5}
\ln M_{\text{bias}} (z) = B_{M,0} + \alpha \ln (1+z)
\end{equation}
and $\sigma _{\ln M}$ the intrinsic scatter in the relation between true and observed mass, 
\begin{equation}\label{eq:s6}
\sigma ^2 _{\ln M} (z)= \sigma ^2 _{\ln M,0} -1 +(1+z)^{2 \beta} \, .
\end{equation}

We stress that in our analysis we assume the bias for the mass estimation and the intrinsic scatter to be redshift dependent. Indeed, while these quantities are usually assumed to be constants, it has been shown that a redshift evolution would be necessary in order to provide a more realistic description of the scaling relations, see e.g. \cite{Salvati:2019zdp} and references therein.

\subsection{Halo Mass Function}

In this investigation, we implement two different formulations for the mass function.
We compare the results of the analysis from \cite{Tinker:2008ff} (T08 hereafter) and \cite{Despali:2015yla} (D16 hereafter), both widely used in the cosmological community. 
We choose to compare these two formulations since 
they represent two approaches when evaluating the mass function.

The analysis in D16 is based on the original formulation from  \cite{Sheth:1999mn} and parametrizes the mass function in terms of 
\begin{equation}\label{eq:d3}
\nu = \left( \dfrac{\delta _c}{\sigma}\right) ^2 \, . 
\end{equation}
As described in D16, $\delta_c$ in Eq.~\ref{eq:d3} is the critical linear theory overdensity $\delta _{\text{lin}}$ required for spherical collapse, divided by the growth factor, with $\delta _{\text{lin}}$ being
\begin{equation}
\delta _{\text{lin}} \simeq \dfrac{3}{20} (12 \pi)^{2/3} \left[ 1+0.0123\,  \log{\Omega(z)}\right] \,. 
\end{equation}
The $\sigma$ quantity in Eq.~\ref{eq:d3} is the standard deviation of density perturbations in a sphere of radius $R = (3M/4 \pi \rho _0)^{1/3}$, defined in linear regime as
\begin{equation}\label{eq:sigma}
\sigma^2 = \dfrac{1}{2 \pi ^2} \int dk \, k^2 P(k,z) \, |W(kR)|^2 \, ,
\end{equation}
\noindent where $W(kR)$ is the window function of a spherical top-hat of radius $R$. 
The mass function then reads
\begin{equation}\label{eq:d1}
\dfrac{dn}{dM} = \nu f(\nu) \dfrac{2 \rho _0}{M} \dfrac{d \ln \sigma ^{-1}}{dM} \, ,
\end{equation}
with
\begin{equation}\label{eq:d2} 
\nu f(\nu) = A \left[ 1 + \left( \dfrac{1}{a\nu} \right)^{p} \right] \left( \dfrac{a \nu }{2 \pi} \right) ^{1/2} \exp{\left(- \dfrac{a \nu}{2} \right)} \,. 
\end{equation}
In order to obtain the coefficients $A$, $a$ and $p$ at $\Delta _c =200$, we follow D16 and adopt the following definitions
\begin{eqnarray}
a &=& 0.4332\,  x^2 + 0.2263 \, x + 0.7665 \\
p &=& -0.1151 \, x^2 + 0.2554 \, x + 0.2488 \\
A &=& -0.1362 \, x + 0.3292  \, 
\end{eqnarray}
where $x$ is defined as $x = \log (\Delta_c/\Delta _{\text{vir}})$. \\

The analysis in T08 formulates the mass function in terms of $\sigma$ (as defined in Eq.~\ref{eq:sigma}). The mass function then reads
\begin{equation}\label{eq:t1}
\dfrac{dn}{dM} = f(\sigma) \dfrac{\rho_0}{M} \dfrac{d \ln \sigma ^{-1}}{dM} \, ,
\end{equation}
with
\begin{equation}\label{eq:t2}
f(\sigma) = A(z) \left[ \left( \dfrac{\sigma}{b(z)} \right)^{-a(z)} +1 \right] \exp{\left( -\dfrac{c(z)}{\sigma^2} \right)} \, .
\end{equation}
The coefficients in Eq.~\ref{eq:t1} are defined as
\begin{eqnarray}
A(z) &=& A_0 (1+z) ^{-0.14}  \label{eq:t3.1}\\
a(z) &=& a_0 (1+z)^{-0.06}  \label{eq:t3.2}\\
b(z) &=& b_0 (1+z)^{-\alpha}  \label{eq:t3.3}\\
\log \alpha (\Delta_c) &=& -\left[ \dfrac{0.75}{\log (\Delta_c /75)} \right] ^{1.2}  \label{eq:t3.4}
\end{eqnarray}
where $A_0$, $a_0$ and $b_0$ are evaluated at redshift $z=0$ for $\Delta_c = 200$.\\

We choose these two formulations also because they both adopt the spherical overdensity algorithm to identify halos. 
To conclude, we note that these formulations provide consistent results, within $10\%$, only in the intermediate mass range and in the redshift range up to $z\leq 1.25$, when considering $\Delta _c = 200$, as discussed in D16. 

\subsection{Characteristics of the forecasted experiments}\label{sec:exp}

In this analysis, we consider galaxy clusters detected through future optical and near-IR galaxy surveys. In order to characterise these surveys and build the mock cluster catalogs, we rely on the observed field of view and the covered redshift range.
For the cluster selection, we consider a minimum mass threshold as a function of redshift.

In details, we provide results mimicking the observational strategy for three future experiments, that will provide galaxy cluster catalogs up to high redshift and low mass. We simulate observations for a Euclid-like \citep{2011arXiv1110.3193L}, a LSST-like \citep{2009arXiv0912.0201L} and a WFIRST-like \citep{2015arXiv150303757S} survey. 

\begin{figure*}[!h]
  \centering
  \subfigure{\includegraphics[scale=0.4]{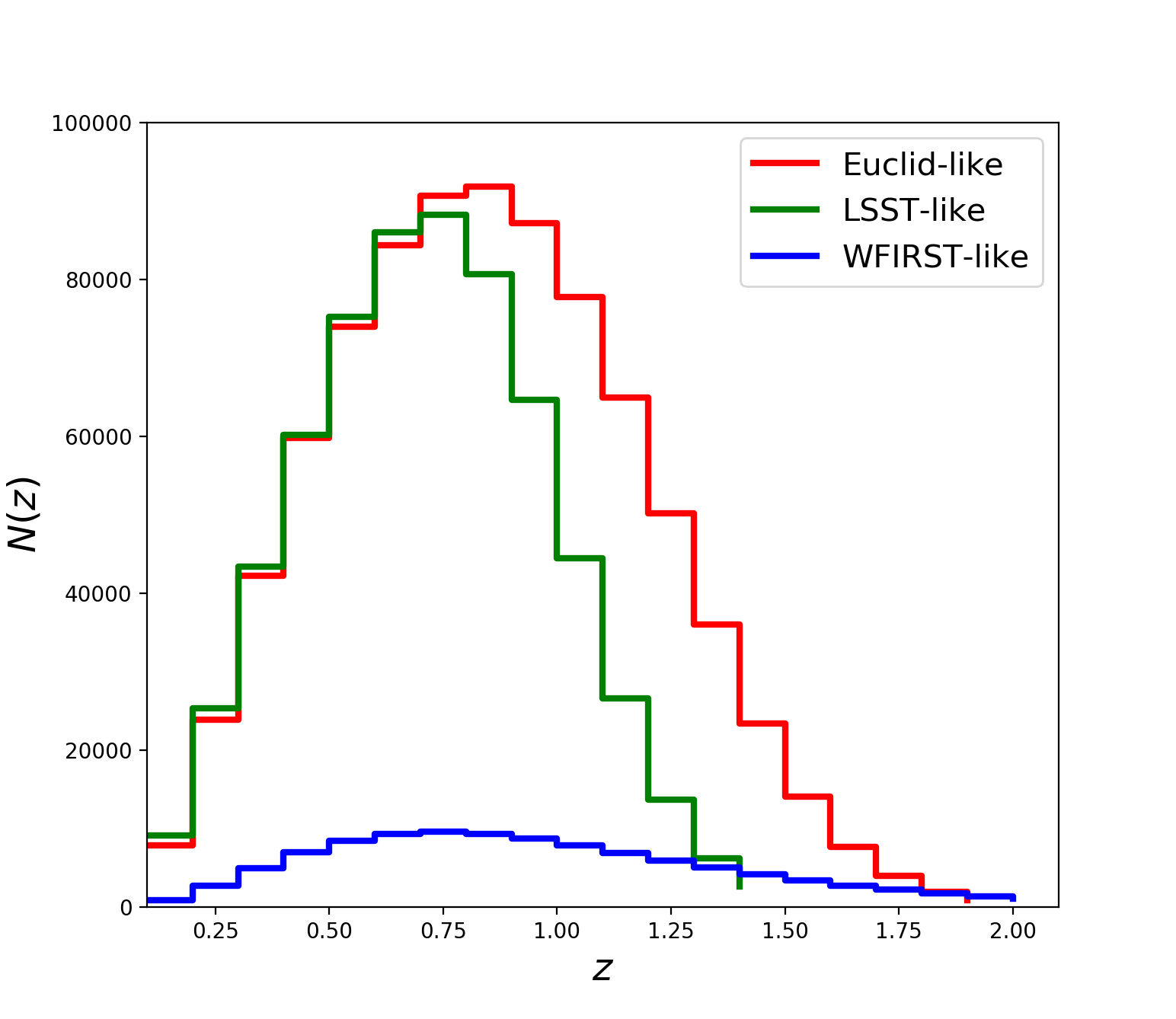}}\,
  \subfigure{\includegraphics[scale=0.4]{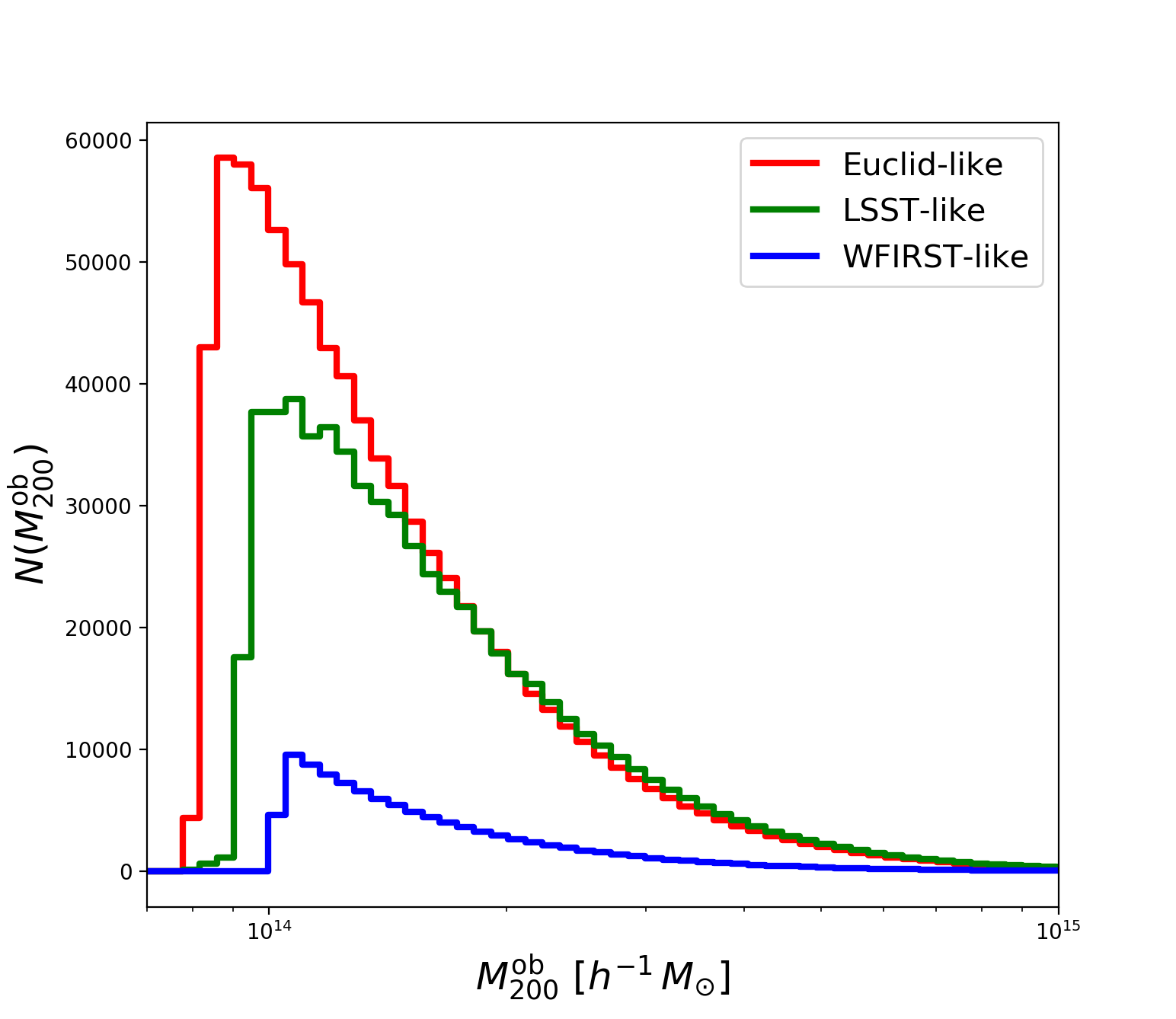}}
  \caption{\footnotesize{Simulated cluster number counts as function of redshift (left) and observed mass (right) for the three experimental setup: Euclid-like (red), LSST-like (green), WFIRST-like (blue).}}
  \label{fig:Nzm}
\end{figure*}

In order to simulate the detected clusters for the Euclid-like and LSST-like surveys, 
we follow the recipe in \cite{Ascaso:2016qlg}. In particular, for the Euclid-like experiment we assume a sky coverage of $15000 \, \text{deg}^2$ and the redshift range $z = [0.1,1.9]$. The mock dataset is characterised by a median redshift $z_{\text{med}} = 0.81$ and a median mass $M^{\text{ob}}_{200,\text{med}} = 1.53 \cdot 10^{14} \, \, [h^{-1} \, M_{\odot}].$ For the LSST-like, we consider a sky coverage of $18000 \, \text{deg}^2$ and the redshift range $z = [0.1,1.4]$. The mock dataset is characterised by a median redshift $z_{\text{med}} = 0.67$ and a median mass $M^{\text{ob}}_{200,\text{med}} = 1.79 \cdot 10^{14} \, \, [h^{-1} \, M_{\odot}].$ In both cases, for the selection function we follow the analysis in \cite{Ascaso:2016qlg}. For the WFIRST-like experiment we follow the recipe in \cite{Gehrels:2014spa} and consider a sky coverage of $2400 \, \text{deg}^2$, with the redshift range $z = [0.1,2.0]$. We assume a cut in mass  for the selection function, $M \ge 10^{14} [M_{\odot} h^{-1}]$. The mock dataset is characterised by a median redshift $z_{\text{med}} = 0.90$ and a median mass $M^{\text{ob}}_{200,\text{med}} = 1.75 \cdot 10^{14} \, \, [h^{-1} \, M_{\odot}].$
We report in Fig.~\ref{fig:Nzm} the simulated cluster number counts as function of redshift, $N(z)$, and observed mass, $N(M^{\text{ob}}_{200})$, for the three experiments.

We stress that when simulating the cluster catalogs, we always adopt the mass function formulation from T08.

\subsection{Analysis}
We adopt a Monte Carlo Markov Chains (MCMC) approach in the forecasts analysis. As MCMC sampler, we use the publicly available package \texttt{cosmomc} \citep{Lewis:2002ah}, which relies on a convergence diagnostic based on Gelman and Rubin statistics. We sample at the same time on the cosmological and scaling relation parameters. 

For the cosmological model, we first assume the $\Lambda$CDM scenario. We vary the six standard parameters: the baryon and CDM densities, $\Omega_b$ and $\Omega_c$; the ratio of the sound horizon to the angular diameter distance at decoupling $\theta$; the scalar spectral index, $n_s$; the overall normalization of the spectrum, $A_s$ at $k = 0.05 \, \text{Mpc}^{-1}$; and reionization optical depth $\tau$. 
When presenting the results, we focus on the parameters describing the matter distribution in the Universe, to which galaxy clusters are more sensitive, i.e. the total matter density $\Omega_m$ and the standard deviation of density perturbations, defined in Eq.~\eqref{eq:sigma}, evaluated at radius $R = 8 \, \text{Mpc} \, h^{-1}$, $\sigma_8$.

We then consider the scenario where the equation of state (EoS hereafter) for dark energy is varying with time. We adopt the parametrisation from \cite{CHEVALLIER2001} and \cite{Linder2003}
\begin{equation}
w = w_0 +(1-a) w_a \, .
\end{equation}
We recall that galaxy cluster number counts alone are not able to constrain the entire set of cosmological parameters. For this reason, we adopt Gaussian priors from the latest Planck release \citep{Aghanim:2018eyx} on the baryon density $\Omega_b h^2$ and the optical depth $\tau$.

The scaling relation parameters are defined in Eqs.~\ref{eq:s5} and \ref{eq:s6} and we adopt the following fiducial values: $B_{M,0} = 0$, $\alpha = 0$, $ \sigma _{\ln M} = 0.2$ and $\beta = 0.125$.
In order to test the impact of the scaling relation calibration on the cosmological results, we consider three different cases where scaling relation parameters are known with an accuracy of $1\% $,  $5\% $ and $10\% $. The adopted values are reported in Tab.~\ref{tab:sr_error}.

\begin{table}[!h]
\begin{center}
\scalebox{0.8}{
\begin{tabular}{c|c|c|c}
\hline
\hline
\morehorsp
Parameter  & $1\% $ & $5\% $ & $10\% $ \\
\hline
\morehorsp
$ B_{M,0}$ &$0.0 \pm 0.001  $ &$0.0 \pm 0.005 $ &$0.0 \pm 0.01 $ \\
\hline
\morehorsp
$\alpha$ &$ 0.0 \pm 0.002$ &$ 0.0 \pm 0.01$ &$0.0 \pm 0.02 $ \\
\hline
\morehorsp
 $\sigma _{\ln M}$ &$0.2 \pm 0.002 $ &$0.2 \pm 0.01 $ &$0.2 \pm 0.02 $  \\ 
\hline
\morehorsp
$\beta$ &$ 0.125 \pm 0.00125 $ &$0.125 \pm 0.00625 $ &$0.125 \pm 0.0125 $  \\ 
\hline
\end{tabular}}
\caption{\footnotesize{Priors on scaling relation parameters applied in the analysis.}}
\label{tab:sr_error}
\end{center}
\end{table}

We compare the results for the three simulated experiments, labelled as "Euclid-like", "LSST-like" and "WFIRST-like". Furthermore, we compare the effect of the implementation of two mass functions T08 and D16 in the pipeline. For this latter analysis, we choose as a baseline the case where scaling relation parameters are known with a $5\%$ accuracy.

\section{Results}\label{sec:results}

In this section we report our results. We focus on how different accuracy for the scaling relations, formulations for the mass function 
and observation strategy
affect the estimation of the cosmological parameters. For the latter, 
as discussed in the previous section, we focus on the results for the matter density $\Omega_m$ and the standard deviation of density perturbations $\sigma _8$.

We start from the $\Lambda$CDM scenario and then discuss an extension of the standard model, with a varying equation of state for dark energy.

\subsection{$\Lambda$CDM}

\begin{figure*}[!h]
  \centering
  \subfigure{\includegraphics[scale=0.55]{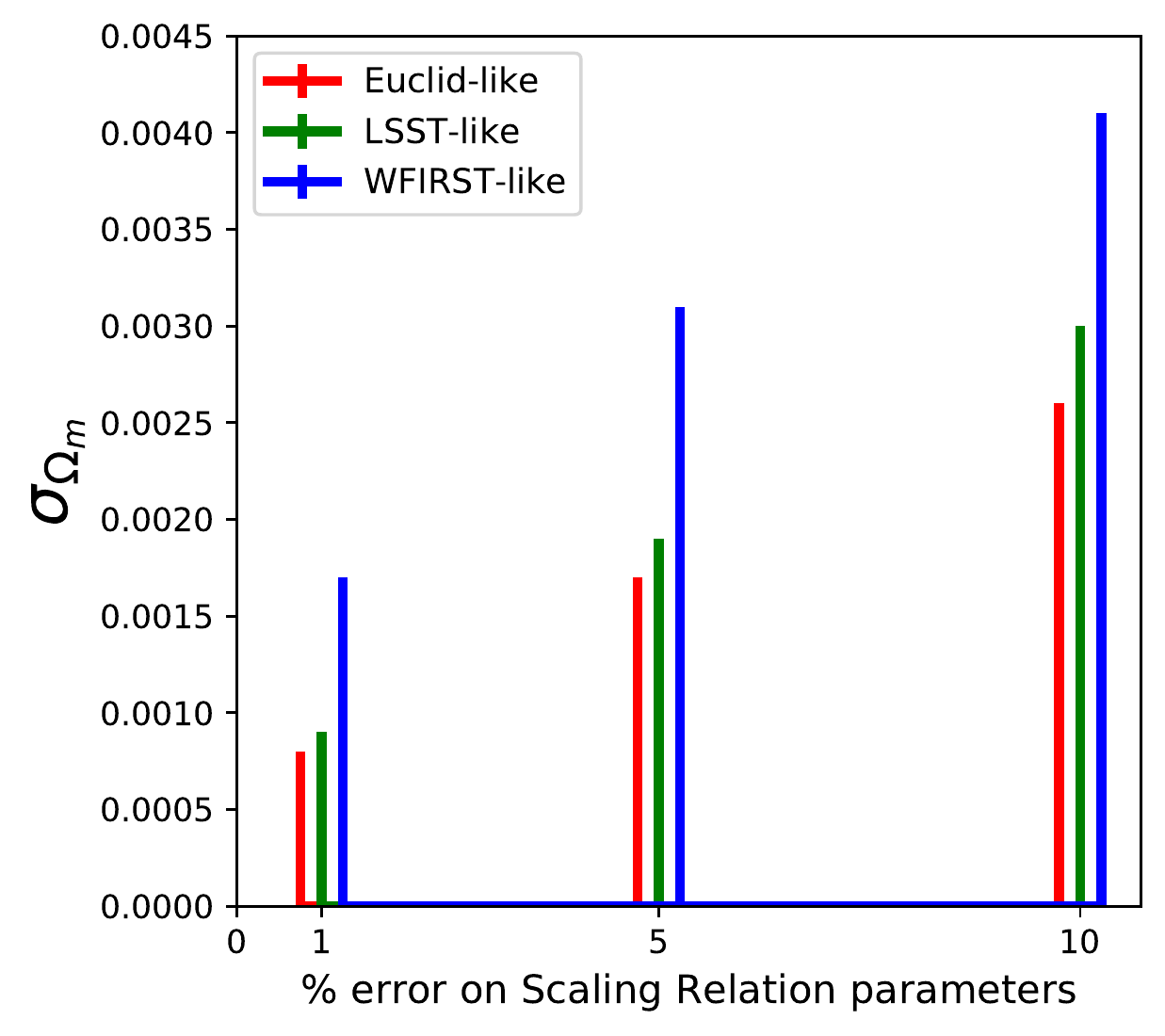}}\,
  \subfigure{\includegraphics[scale=0.55]{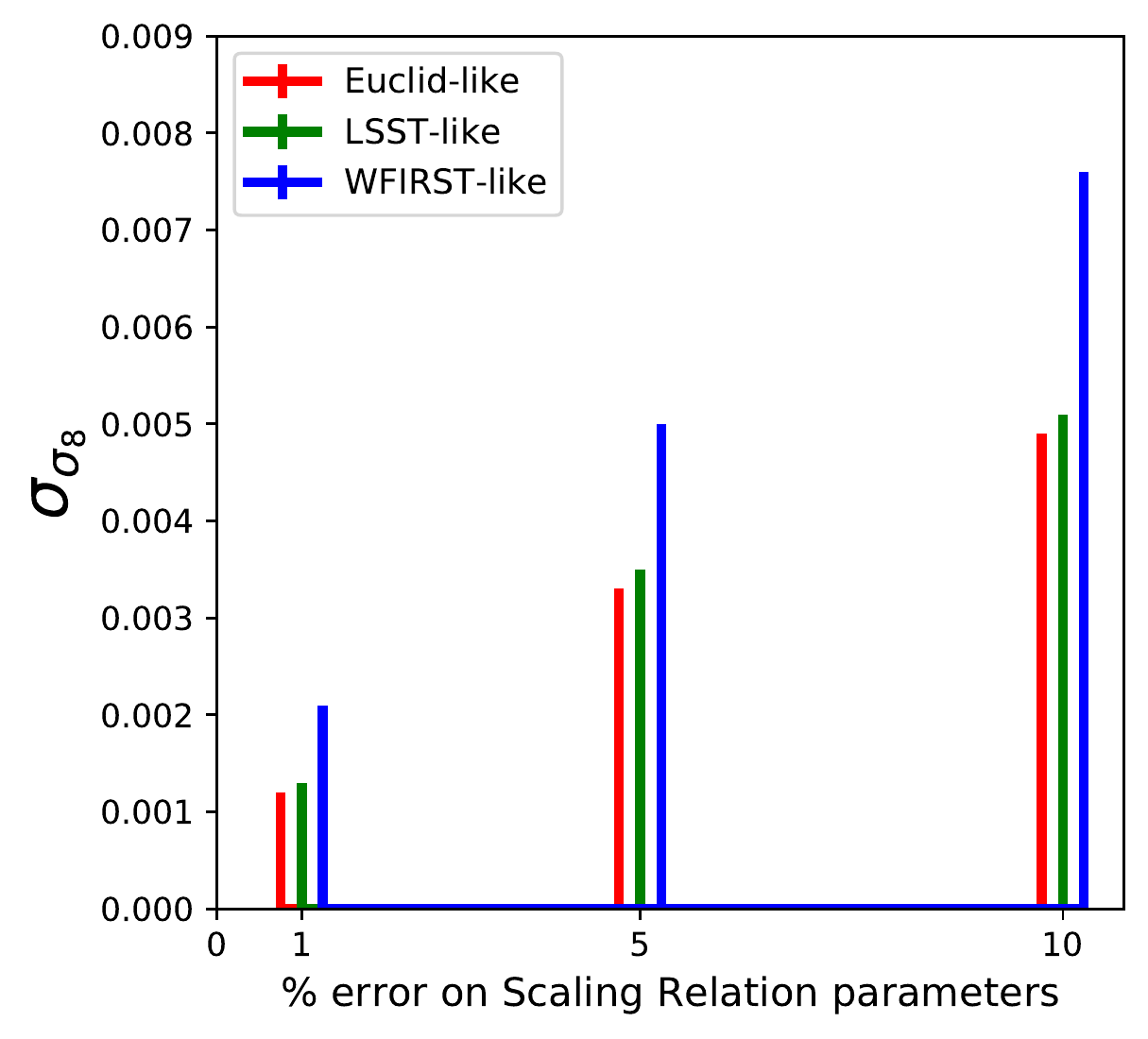}}
  \caption{\footnotesize{Error on the cosmological parameters $\Omega_m$ and $\sigma_8$ for different accuracies on the scaling relation parameters ($1\%$, $5\%$ and $10\%$). We report results for the Euclid-like (red), the LSST-like (green) and the WFIRST-like (blue) experiments.}}
  \label{fig:sigma_LCDM}
\end{figure*}

\begin{figure*}[!h]
  \centering
  \subfigure{\includegraphics[scale=0.55]{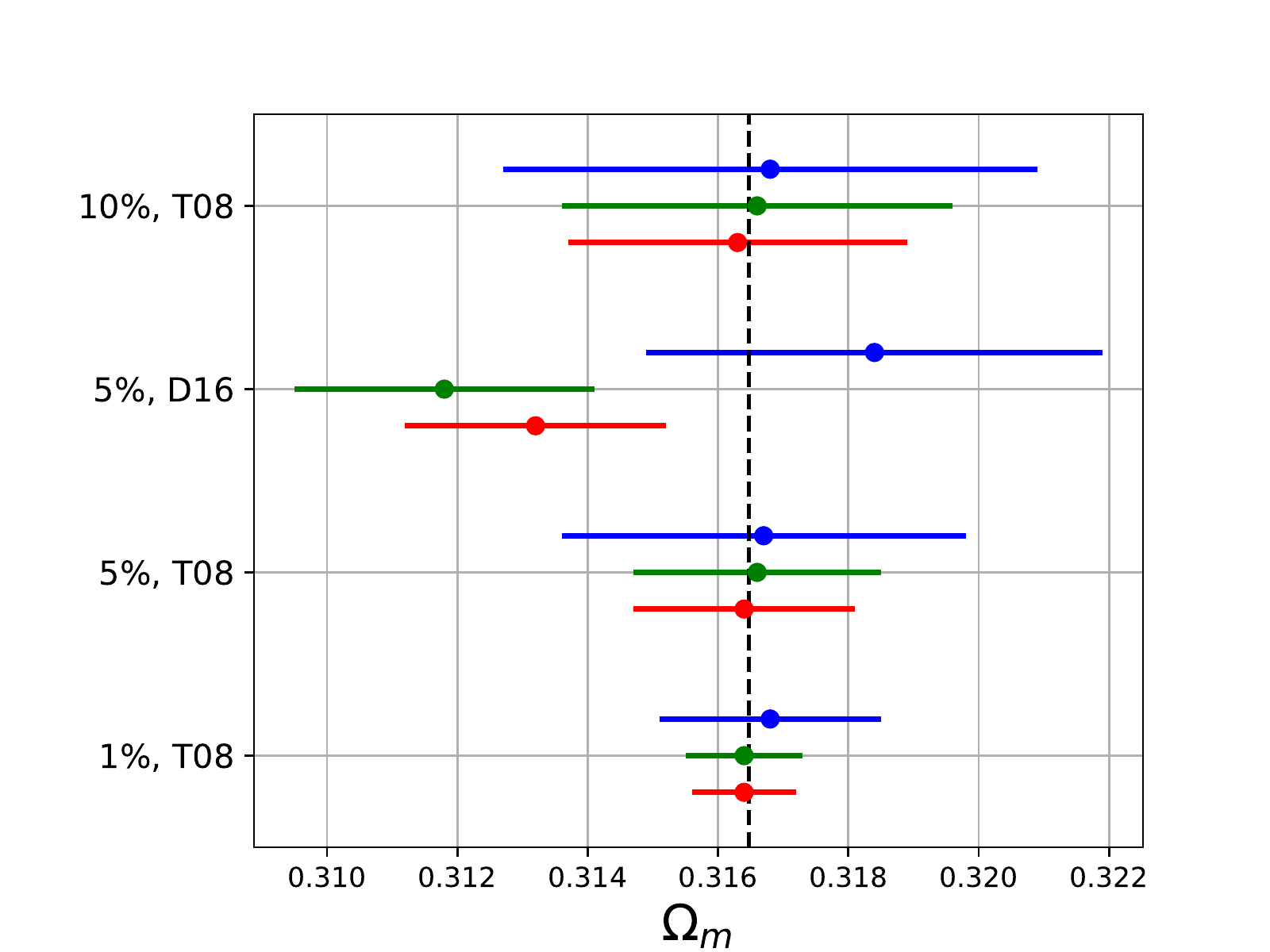}}\,
  \subfigure{\includegraphics[scale=0.55]{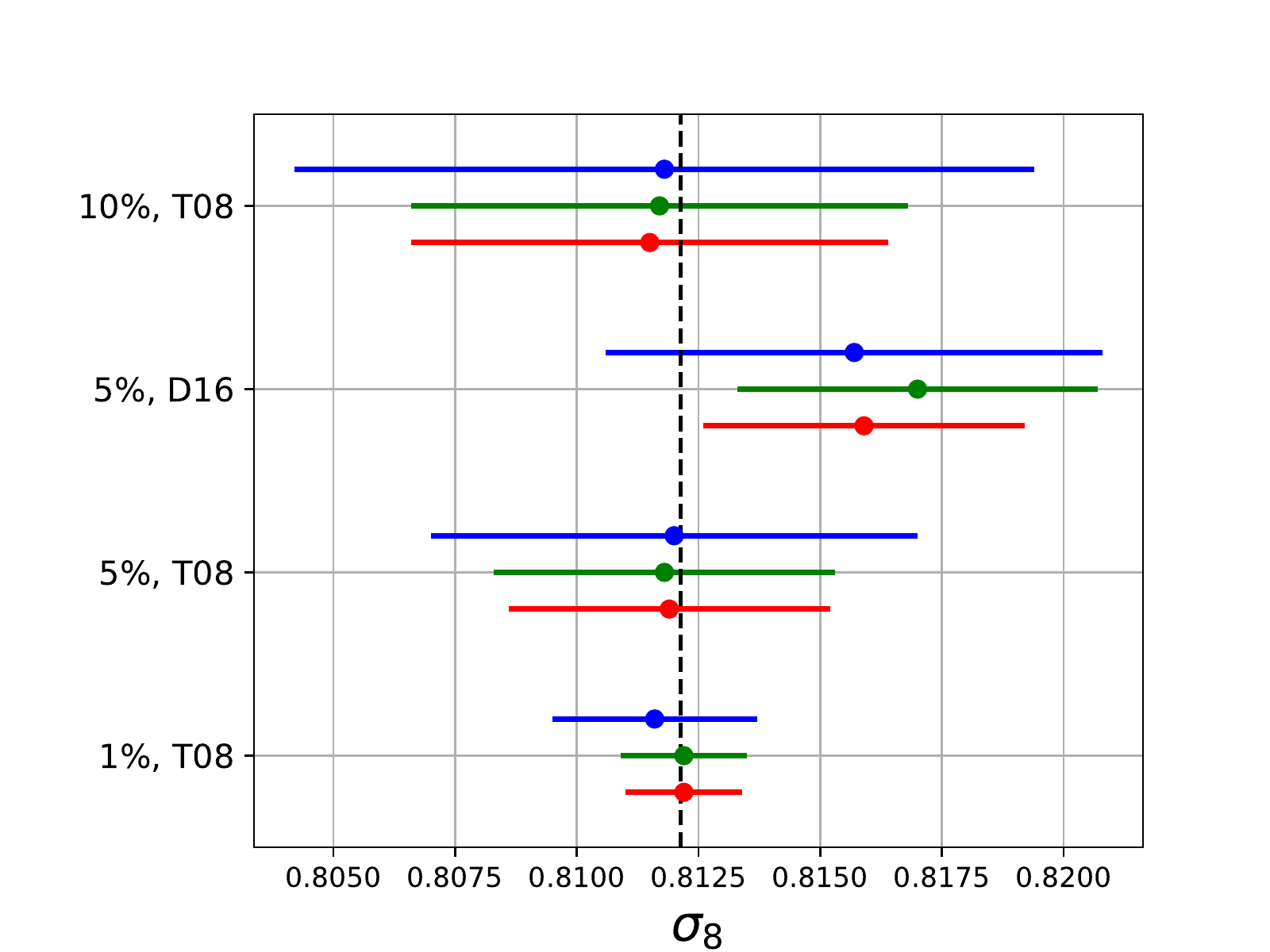}}
  \caption{\footnotesize{Values of $\Omega_m$ and $\sigma_8$ with $1 \, \sigma$ errorbars. We report results for the Euclid-like (red), the LSST-like (green) and WFIRST-like (blue) experiments, for different scaling relation parameter accuracies and the two mass function formulations. The black vertical dashed line represents the input value adopted for the mock data.}}
  \label{fig:whisker_lcdm}
\end{figure*}

We start discussing the effect of the different accuracy on the scaling relation parameters. We consider three different scenarios, where scaling relations parameters are known with an accuracy of $10\%$, $5\%$ and $1\%$. 
We report the $68\%$ confidence level (c.l. from now on) constraints for the scaling relation and cosmological parameters in Tab.~\ref{tab:LCDM}. 
In Figs.~\ref{fig:sigma_LCDM} and \ref{fig:whisker_lcdm} we report the errors on the cosmological parameters and the $1\, \sigma$ constraints, for the different accuracies on the scaling relations and for the three experiment configurations.
As expected, the improvement in the accuracy  leads to an increasing constraining power on the cosmological parameters, given the degeneracies between these parameters. 

In order to discuss how the different experiment characterisations might affect the final results, 
we choose as a baseline the case where scaling relation parameters are known with a $5\%$ accuracy.
 
We show the comparison between the different experiments, for the cosmological and scaling relation parameters, in the triangular plot in Fig.~\ref{fig:5SR_LCDM_allexp}, in Appendix \ref{sec:app1}.
On the one hand, the Euclid-like and LSST-like experiments provide tight, consistent constraints. As described in section \ref{sec:exp}, the two simulated experiments are indeed characterised by a similar sky coverage, while having a different redshift range and selection function. We stress that the tighter constraints on the cosmological parameters are also due to the better shaping of the degeneracy with scaling relation parameters.
On the other hand, the WFIRST-like experiment provides wider constraints on the cosmological parameters. This experiment is simulated with a lower sky coverage, while spanning up to redshift $z=2$ with a flat selection function. 
From these results we therefore deduce that the precision on the constraints on cosmological parameters is affected also by the experiment sky coverage.

\begin{figure*}[!h]
  \centering
  \subfigure{\includegraphics[scale=0.315]{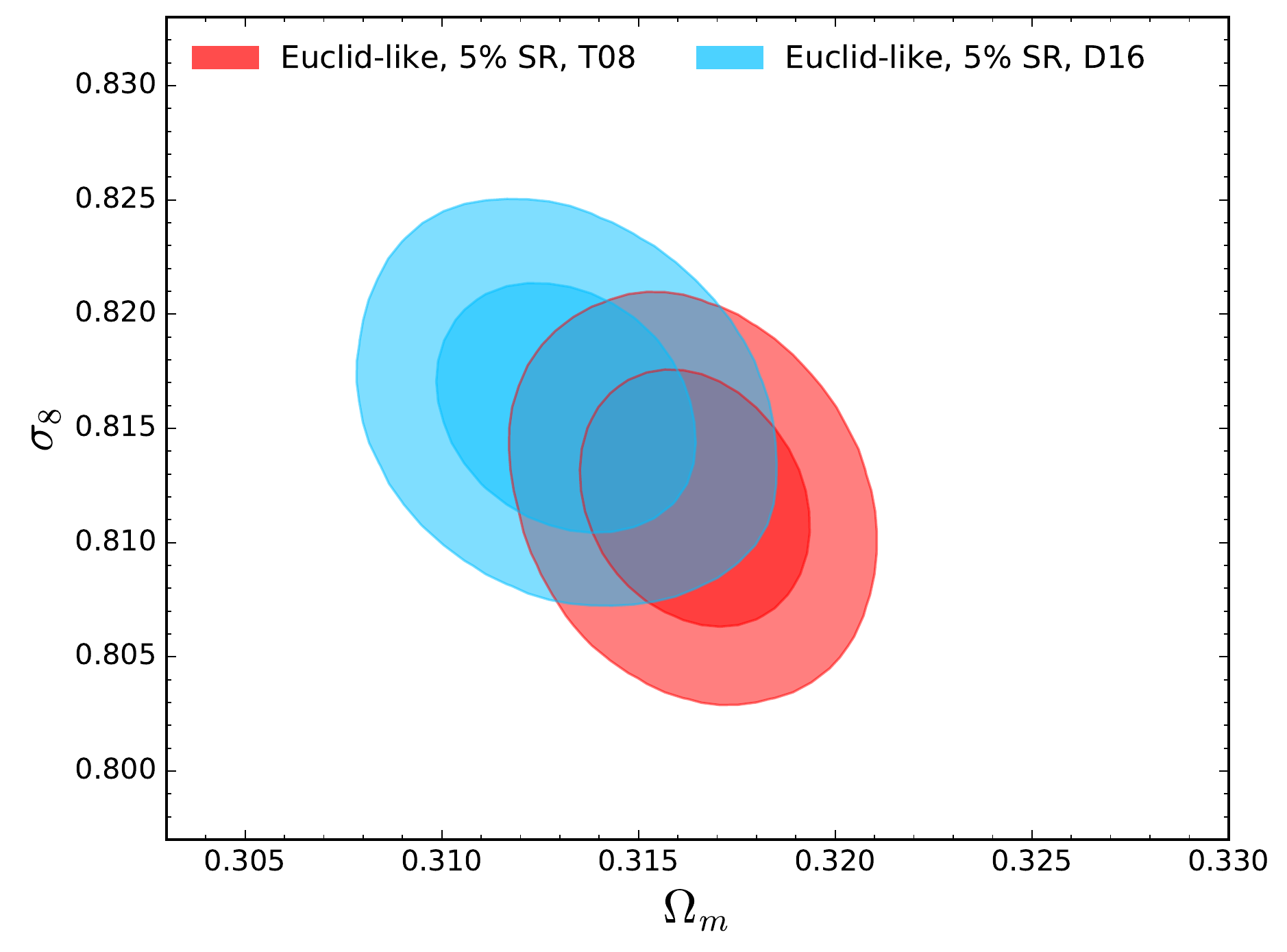}}\,
  \subfigure{\includegraphics[scale=0.315]{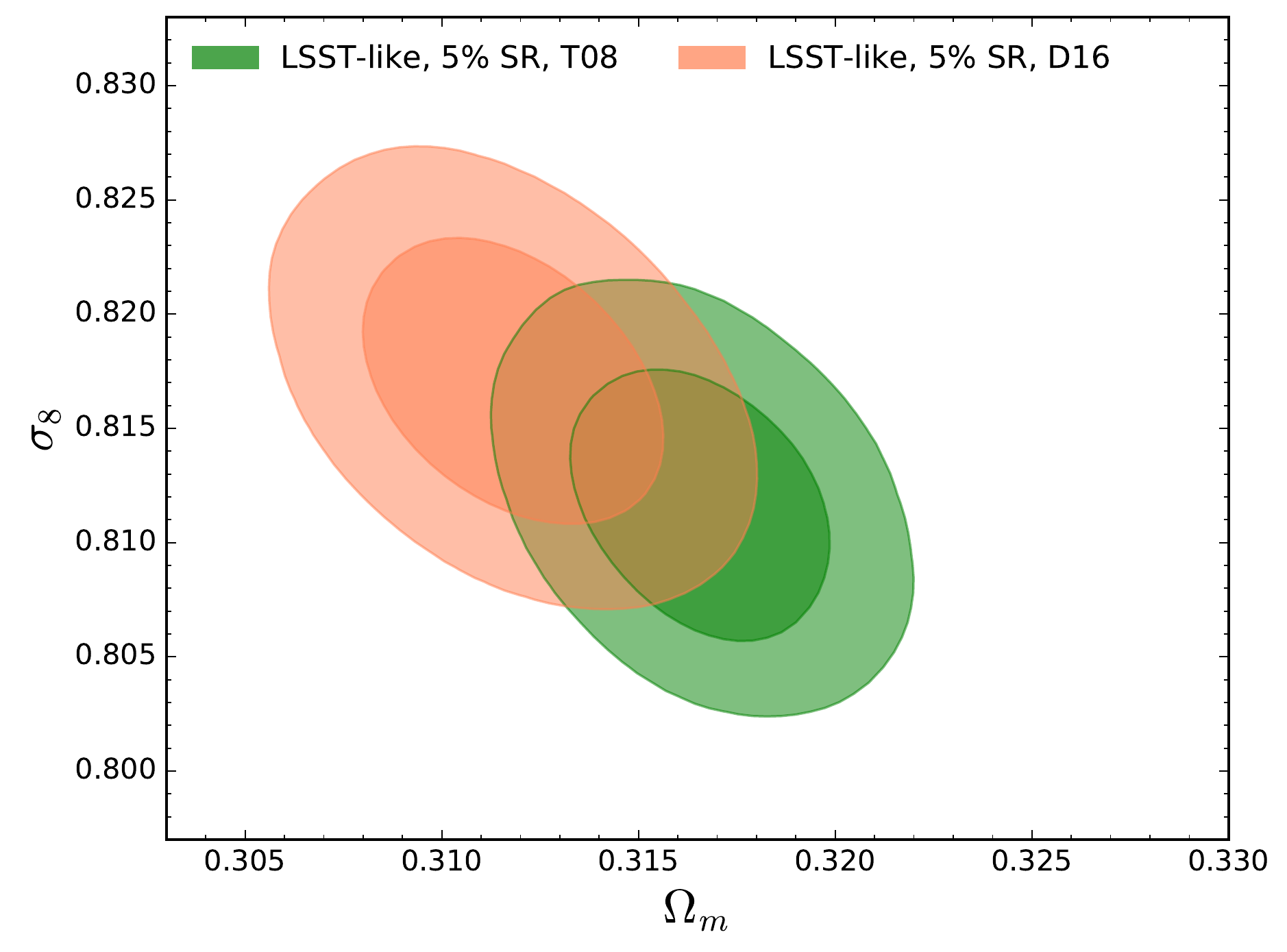}}\,
  \subfigure{\includegraphics[scale=0.315]{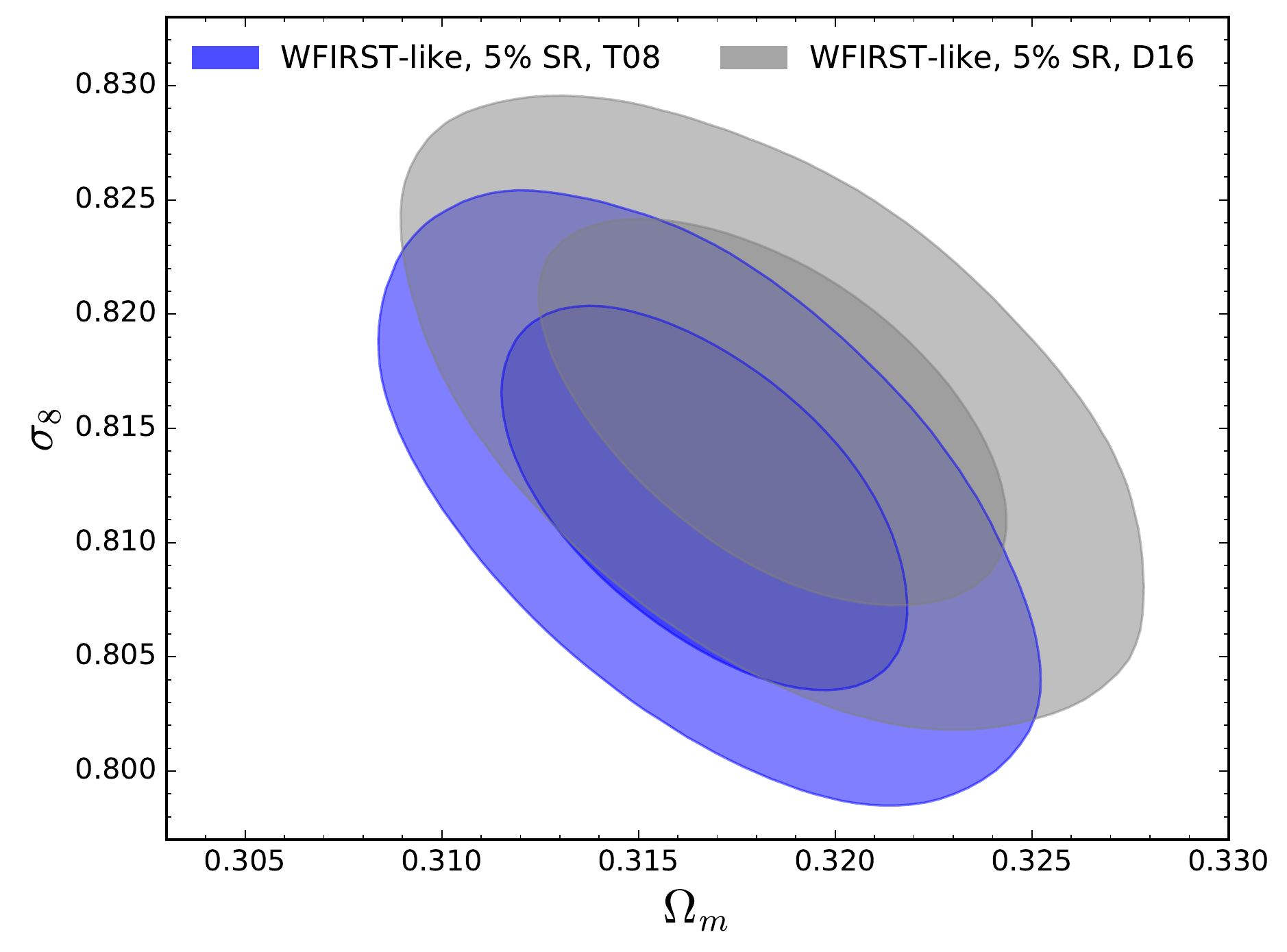}}
  \caption{\footnotesize{Two-dimensional probability distributions for $(\Omega_m,\sigma_8)$. We show results for a $5\%$ accuracy on the scaling relation parameters, considering two evaluations of the mass function, T08 and D16. The plots correspond to the Euclid-like survey (left panel), the LSST-like survey (middle panel) and the WFIRST-like survey (right panel).}}
  \label{fig:MF_comp_all}
\end{figure*}

We conclude this section discussing the effects of the different evaluations for the mass function. 
We consider again as the baseline the case with a $5\%$ accuracy on the scaling relation parameters and we compare results obtained using the mass function evaluations from T08 and D16. The $68\%$ c.l. results for the latter are also reported in Tab.~\ref{tab:LCDM} and the $1 \, \sigma$ constraints for $\Omega_m$ and $\sigma_8$ are shown in Fig.~\ref{fig:whisker_lcdm}.
In Fig.~\ref{fig:MF_comp_all} we show the two-dimensional probability distributions for $(\Omega_m, \sigma_8)$ for the three simulated experiments, comparing results for the two mass functions.

We recall here that when producing the simulated cluster catalogs, we adopt the T08 formulation for the mass function.
When comparing the effect of the mass function formulations in our analysis, on the one hand we confirm that we reproduce the input values for the cosmological parameters when using T08, as expected.
On the other hand, the use of D16 introduces biases and shifts in the final results.

We stress therefore that the two mass function implementations do not recover the same cosmological parameter constraints.
In particular, the impact of the choice of the mass function can be mainly seen on the results from the Euclid- and LSST-like experiments, producing a shift along the $(\Omega_m,\sigma_8)$ degeneracy line up to $1.6 \, \sigma$. Indeed, the lower precision of the WFIRST-like experiment provides wider errors on the parameter constraints, therefore not showing the difference between results for the two mass function. 

\begin{figure*}[!ht]
\centering
\includegraphics[scale=0.27]{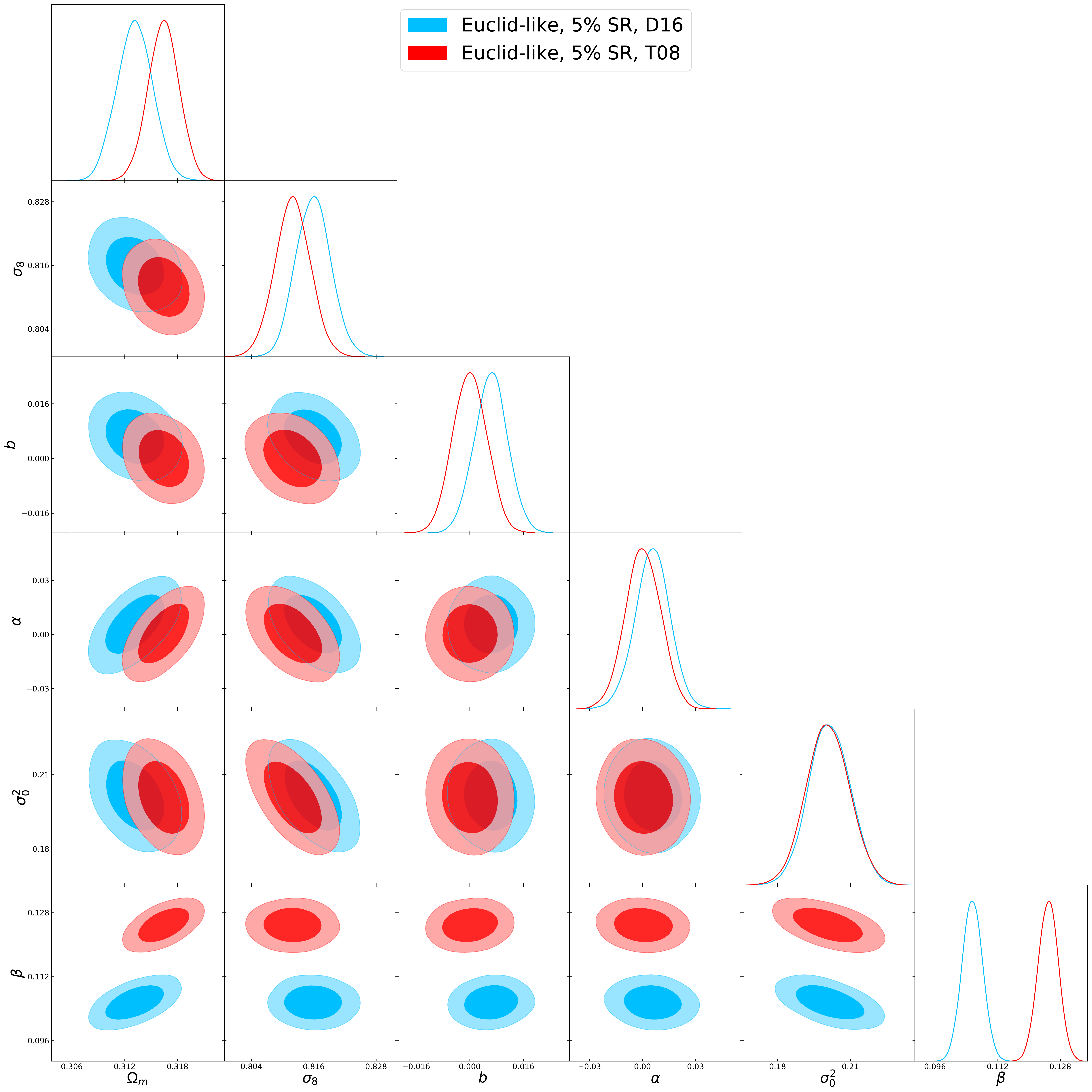}
\caption{\footnotesize{Constraints on cosmological and scaling relation parameters, in the $\Lambda$CDM scenario, for the comparison between the two mass function formulations for the Euclid experiment. We show results for the $5\%$ accuracy on the scaling relation parameters, for T08 mass function (red) and D16 mass function (light blue).}}
\label{fig:MF_LCDM_Euclid}
\end{figure*}

Given the consistency between the Euclid-like and LSST-like experiments, we focus on the first one to further discuss these results. 
In the triangular plot in Fig.~\ref{fig:MF_LCDM_Euclid} we show the constraints on cosmological and scaling relation parameters for the Euclid-like mission. 
From these results, we stress that the change from T08 to D16 mass function in the analysis provides also a shift on the scaling relation parameters. In particular, the largest effect can be seen on the $\beta$ parameter. This quantity parametrises the redshift evolution of the scatter of the scaling relations and shows a shift $> 7 \, \sigma$ towards lower values when adopting D16 evaluation.
This shift may represent a general different redshift evolution for the two mass functions, that is indeed driving the constraints on $(\Omega_m,\sigma_8)$.

\begin{figure*}[!ht]
\centering
\includegraphics[scale=0.3]{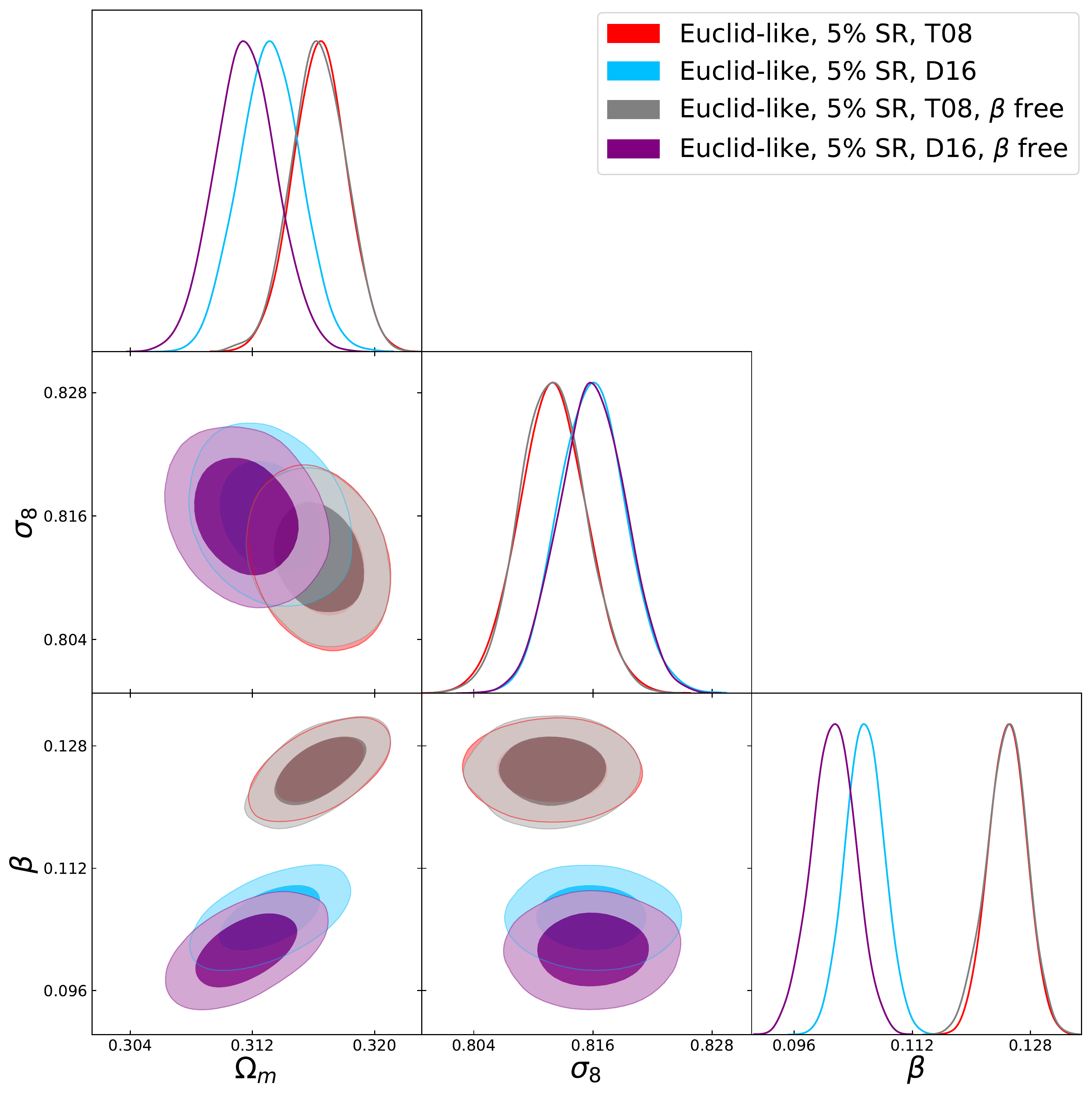}
\caption{\footnotesize{Constraints on cosmological and $\beta$ parameters for the Euclid-like experiments, in the $\Lambda$CDM scenario. We assume the scaling relation parameters to be known with a $5\%$ accuracy and compare the results for T08 and D16 mass functions, when considering (red and light blue contours) or not (grey and purple contours) the Gaussian prior on the scaling relation scatter $\beta$.}}
\label{fig:beta_free}
\end{figure*}

In order to better understand this behaviour, as a further test we check the results when letting the $\beta$ parameter unconstrained, i.e. not considering the Gaussian prior $\beta = 0.125 \pm 0.00625$, defined in Tab.~\ref{tab:sr_error}. We show the results for $\beta$ and the cosmological parameters $\Omega_m$ and $\sigma_8$ in Fig.~\ref{fig:beta_free}. We stress that in this case the parameter shift is even enhanced when considering D16 mass function, while results remain consistent for T08 mass function. We confirm therefore that we are mimicking a different redshift evolution for the two mass functions. 

Indeed, the shift of the scaling relation scatter with respect to redshift implies a change in the total cluster number counts $dN/dz$. From previous results (and from discussion in D16), D16 mass function predicts more clusters at higher z. Therefore,  the shift on the scaling relation scatter, and in particular having $\beta_{\text{D16}} < \beta_{\text{T08}}$, is compensating for this effect.

\begin{table*}[!h]
\begin{center}
\scalebox{0.8}{
\begin{tabular}{c||c|c||c|c|c|c}
\hline
\hline
\morehorsp
Experiment  & $\Omega_m$ & $\sigma_8 $ & $B_{M,0} $& $\alpha$ &$\sigma _{\ln M}$ & $\beta$\\
\hline
\hline
\morehorsp
Euclid-like, $1\%$ SR, T08 & $0.3164 \pm 0.0008$ &$ 0.8122 \pm 0.0012$ &$ 0.0 \pm 0.001$ & $ 0.0 \pm 0.0020$ & $ 0.2000 \pm 0.0020$ & $ 0.1250 \pm 0.0011$ \\
\hline
\morehorsp
Euclid-like, $5\%$ SR, T08 &$ 0.3164 \pm 0.0017$ & $ 0.8119 \pm 0.0033$ & $ 0.0 \pm 0.005$ & $ 0.0 \pm 0.010$ & $ 0.2008 \pm 0.0086$ & $ 0.1250 \pm 0.0025$ \\
\hline
\morehorsp
Euclid-like, $5\%$ SR, D16 & $ 0.3132 \pm 0.0020$ & $ 0.8159 \pm 0.0033$ & $ 0.0063 \pm 0.0048$ & $ 0.0056 \pm 0.0094$ & $ 0.2014 \pm 0.0083$ & $ 0.1056 \pm 0.0025$ \\
\hline
\morehorsp
Euclid-like, $10\%$ SR, T08 & $ 0.3163 \pm 0.0026$ & $0.8115 \pm 0.0049 $ & $0.0002 \pm 0.0092 $ & $ 0.001 \pm 0.019 $ & $0.203 \pm 0.014 $ & $ 0.1244 \pm 0.0035 $ \\
\hline
\hline
\morehorsp
LSST-like, $1\%$ SR, T08 & $ 0.3164 \pm 0.0009$& $ 0.8122 \pm 0.0013$ & $ 0.0 \pm 0.001 $ &$ 0.0 \pm 0.002$ & $ 0.200 \pm 0.002$ & $ 0.1250 \pm 0.0012$ \\
\hline
\morehorsp
LSST-like, $5\%$ SR, T08 & $ 0.3166 \pm 0.0019$ & $ 0.8118 \pm 0.0035$ & $ 0.0 \pm 0.0049$ & $ 0.0006 \pm 0.0099$ & $ 0.2006 \pm 0.0086$ & $ 0.1252 \pm 0.0033$ \\
\hline
\morehorsp
LSST-like, $5\%$ SR, D16 &$ 0.3118 \pm 0.0023$ & $ 0.8170 \pm 0.0037$ & $ 0.0027 \pm 0.0049$ & $ 0.0024 \pm 0.0098$ & $ 0.1969 \pm 0.0087$ & $ 0.1099 \pm 0.0034$ \\
\hline
\morehorsp
LSST-like, $10\%$ SR, T08 & $ 0.3166 \pm 0.0030$ & $ 0.8117 \pm 0.0051$ & $ -0.0003 \pm 0.0097$ & $ 0.002 \pm 0.020$ & $ 0.202 \pm 0.014$ & $ 0.1247 \pm 0.0044$ \\
\hline
\hline
\morehorsp
WFIRST-like, $1\%$ SR, T08 & $ 0.3168 _{-0.0018} ^{+0.0016}$ & $ 0.8116 _{-0.0022}^{+0.0020}$ & $ 0.0 \pm 0.001$ & $ 0.0 \pm 0.002$ & $ 0.2001 \pm 0.0019$ & $ 0.1250 \pm 0.0012$ \\
\hline
\morehorsp
WFIRST-like, $5\%$ SR, T08 & $ 0.3167 \pm 0.0031$ & $ 0.8120 \pm 0.0050$ & $ -0.0002 \pm 0.0050$ & $ 0.0 \pm 0.01$ & $ 0.2001 \pm 0.0095$ & $ 0.1254 \pm 0.0039$ \\
\hline
\morehorsp
WFIRST-like, $5\%$ SR, D16 & $ 0.3184 \pm 0.0035$ & $ 0.8157 \pm 0.0051$ & $ 0.0018 \pm 0.0049$ & $ 0.0 \pm 0.010$ & $ 0.1960 \pm 0.0095$ & $ 0.1113 \pm 0.0041$ \\
\hline
\morehorsp
WFIRST-like, $10\%$ SR, T08 & $ 0.3168 \pm 0.0041$ & $ 0.8118 \pm 0.0076$ & $ -0.0003 \pm 0.0098$ & $ 0.0 \pm 0.02$ & $ 0.200 \pm 0.018$ & $ 0.1253 \pm 0.0056$ \\
\hline
\end{tabular}}
\caption{\footnotesize{We report the $68\%$ c.l. for the cosmological and scaling relation parameters in the $\Lambda$CDM scenario. We report results for the Euclid-like, LSST-like and WFIRST-like simulated experiments, at different scaling relation accuracies and for the T08 and D16 mass function formulations.}}
\label{tab:LCDM}
\end{center}
\end{table*}

\subsection{DE EoS}

In this section we report the results when varying the equation of state for dark energy. We adopt the parametrisation $w = w_0 +(1-a) w_a$ .

We follow the same approach as for the standard cosmological scenario and analyse the impact on the cosmological parameter constraints of different accuracies for the scaling relation parameters, different observation strategies and mass function implementations. 

We report the constraints on cosmological and scaling relation parameters obtained for the different configurations in Tab.~\ref{tab:w0wa_cosmo} and ~\ref{tab:w0wa_sr}.

\begin{figure*}[!h]
  \centering
  \subfigure{\includegraphics[scale=0.55]{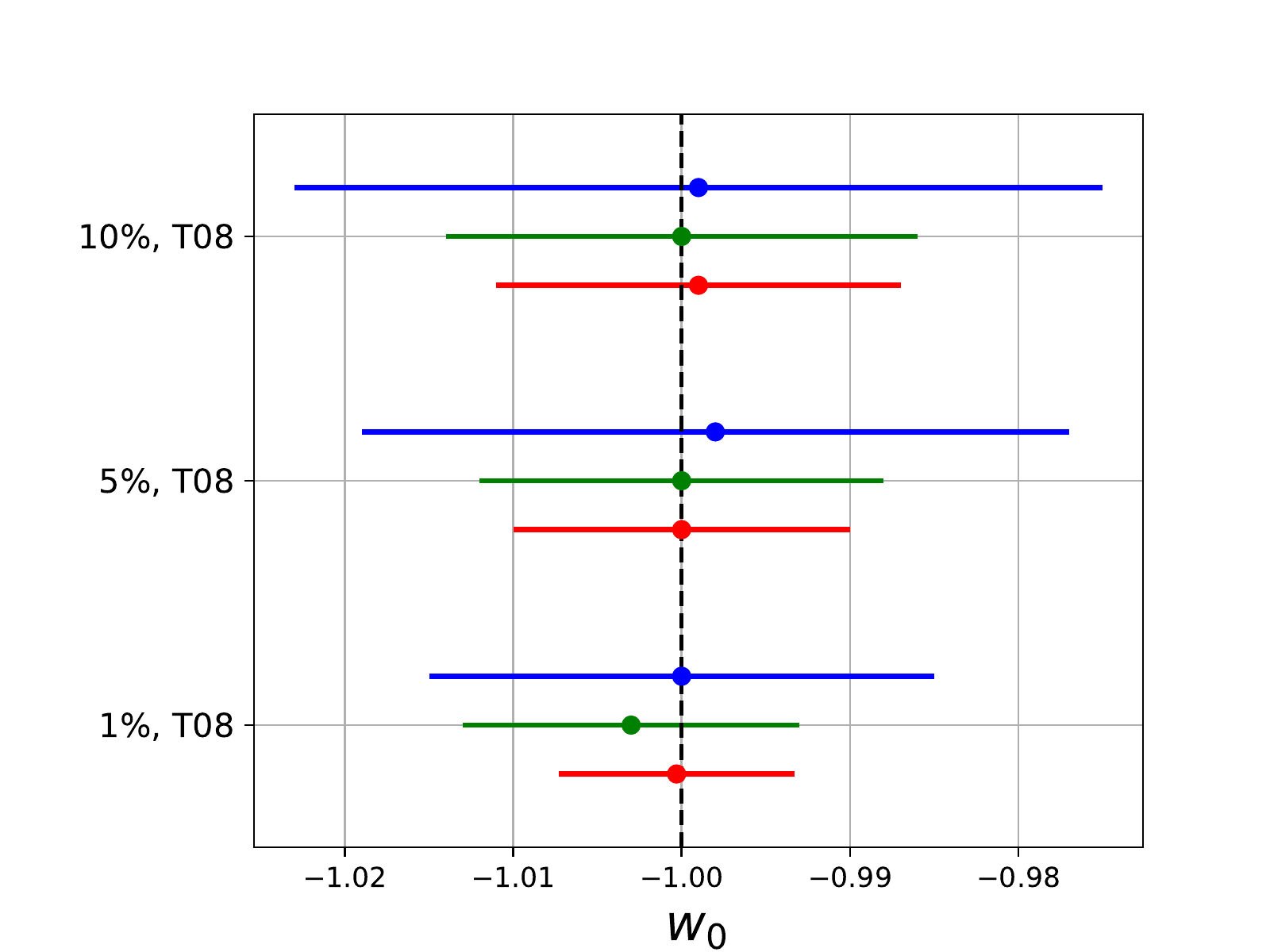}}\,
  \subfigure{\includegraphics[scale=0.55]{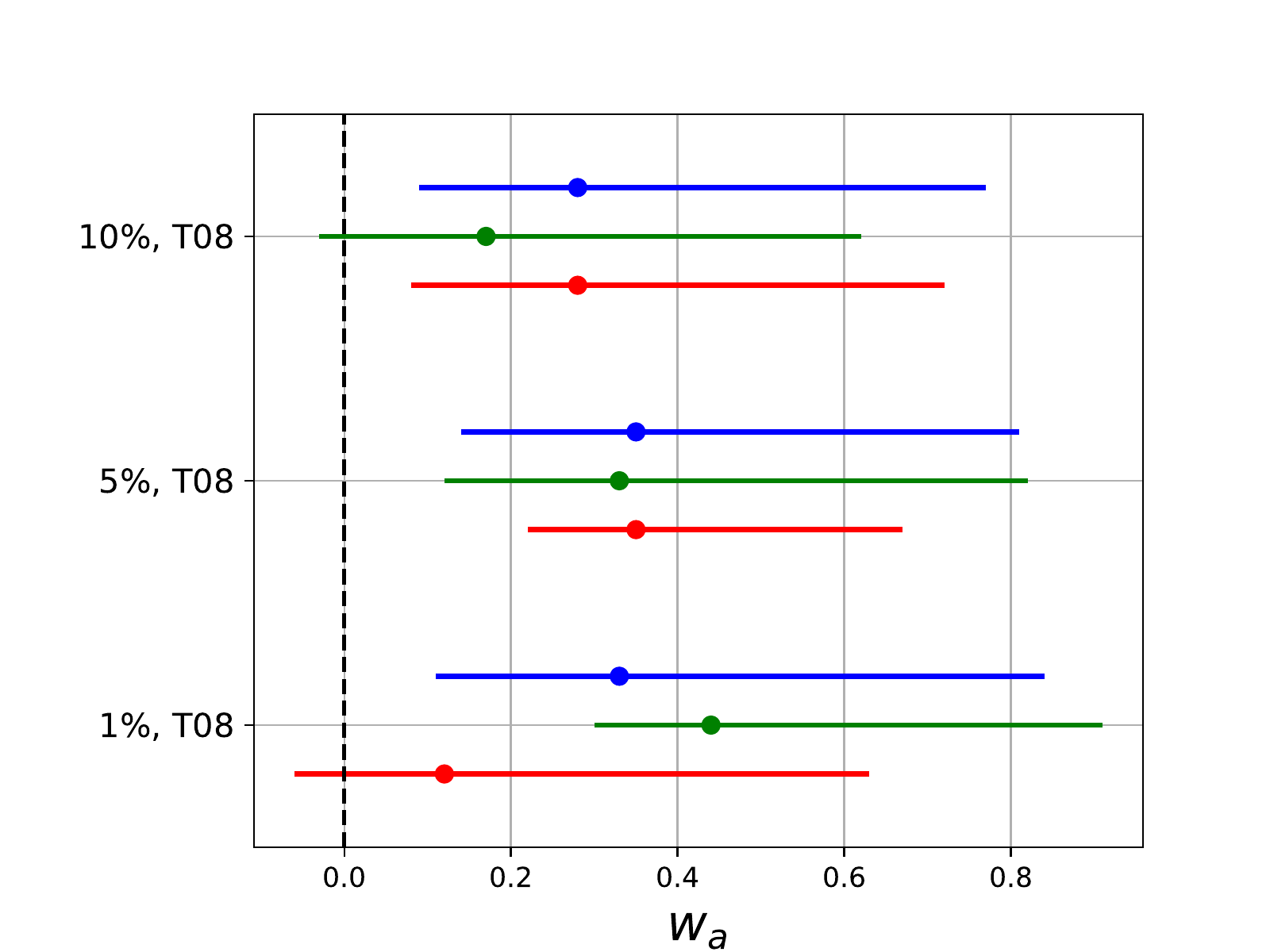}}
  \caption{\footnotesize{Values of $w_0$ and $w_a$ with $1 \, \sigma$ error bars. We report results for the Euclid-like (red), the LSST-like (green) and WFIRST-like (blue) experiments, for different scaling relation parameter accuracies. The black vertical dashed line represents the input value adopted for the mock data.}}
  \label{fig:sigma_w0wa}
\end{figure*}

In Fig.~\ref{fig:sigma_w0wa} we report the values of the cosmological parameters with $1 \, \sigma$ error bars, focusing on $w_0$ and $w_a$. 
In general, as seen in the previous section, the increasing accuracy on scaling relation calibration improves the cosmological results. 
Nevertheless, we note that for the $w_a$ parameter, results remain almost unchanged. This may be due to the fact that cluster number counts alone, without the addition of other cosmological probes, are not able to fully constrain the possible redshift evolution of the EoS for dark energy.

As for the $\Lambda$CDM scenario, the Euclid-like and LSST-like surveys provide consistent results, while the WFIRST-like experiment is still producing wider constraints.

\begin{table*}[!h]
\begin{center}
\scalebox{0.8}{
\begin{tabular}{c||c|c|c|c}
\hline
\hline
\morehorsp
Experiment  & $\Omega_m$ & $\sigma_8 $ & $w_0$ & $w_a$\\
\hline
\morehorsp
Euclid-like, $1\%$ SR, T08 & $0.3164 \pm 0.0013$ & $ 0.8121 \pm 0.0012$ & $ -1.0003 _{-0.0064}^{+0.0072}$ & $ 0.12 _{-0.18}^{+0.51}$ \\
\hline
\morehorsp
Euclid-like, $5\%$ SR, T08 & $ 0.3165 \pm 0.0018$ & $ 0.8122 \pm 0.0033$ & $ -1.000 \pm 0.010$ & $ 0.35 _{-0.13}^{+0.32}$ \\
\hline
\morehorsp
Euclid-like, $5\%$ SR, D16 & $ 0.3066 \pm 0.0020$ & $ 0.8153 \pm 0.0034$ & $ -1.095 \pm 0.012$ & $ 0.13 _{-0.26}^{+0.51}$ \\
\hline
\morehorsp
Euclid-like, $10\%$ SR, T08 & $ 0.3163 \pm 0.0027$ & $ 0.8121 \pm 0.0051$ & $ -0.999 _{-0.011}^{+0.012}$ & $ 0.28 _{-0.20}^{+0.44}$ \\
\hline
\hline
\morehorsp
LSST-like, $1\%$ SR, T08 & $ 0.3161 _{-0.0016} ^{+0.0014}$ & $0.8124 _{-0.0013}^{+0.0012} $ & $ -1.003 \pm 0.010$ & $ 0.44 _{-0.14}^{+0.47}$ \\
\hline
\morehorsp
LSST-like, $5\%$ SR, T08 & $ 0.3167 \pm 0.0020$ & $ 0.8121 _{-0.0039}^{+0.0034}$ & $ -1.000 \pm 0.012$ & $ 0.33 _{-0.21}^{+0.49}$ \\
\hline
\morehorsp
LSST-like, $10\%$ SR, T08 & $ 0.3165 \pm 0.0030$ & $ 0.8118 \pm 0.0050$ & $ -1.000 \pm 0.014$ & $ 0.17 _{-0.20}^{+0.45}$ \\
\hline
\hline
\morehorsp
WFIRST-like, $1\%$ SR, T08 & $ 0.3164 _{-0.0033}^{+0.0037}$ & $ 0.8342 \pm 0.0036$ & $ -1.000 \pm 0.015$ & $ 0.33 _{-0.22}^{+0.51}$ \\
\hline
\morehorsp
WFIRST-like, $5\%$ SR, T08 &$ 0.3169 \pm 0.0038$ & $ 0.8119 \pm 0.0050$ & $ -0.998 \pm 0.021$ & $ 0.35 _{-0.21}^{+0.46}$ \\
\hline
\morehorsp
WFIRST-like, $10\%$ SR, T08 & $ 0.3167 \pm 0.0044$ & $ 0.8122 \pm 0.0079$ & $ -0.999 \pm 0.024$ & $ 0.28 _{-0.19} ^{+0.49}$ \\
\hline
\end{tabular}}
\caption{\footnotesize{We report the $68\%$ c.l. for the cosmological parameters when varying the EoS for dark energy. We report results for the Euclid-like, LSST-like and WFIRST-like simulated experiments, at different scaling relation accuracies. Only for the Euclid-like experiment, we show the comparison between the T08 and D16 mass function formulations.}}
\label{tab:w0wa_cosmo}
\end{center}
\end{table*}

\begin{table*}[!h]
\begin{center}
\scalebox{0.8}{
\begin{tabular}{c||c|c|c|c}
\hline
\hline
\morehorsp
Experiment  & $B_{M,0} $& $\alpha$ &$\sigma _{\ln M}$ & $\beta$\\
\hline
\morehorsp
Euclid-like, $1\%$ SR, T08 & $ 0.0 \pm 0.001$ & $ 0.0 \pm 0.0020$ & $ 0.2000 _{-0.0019}^{+0.0018}$ & $ 0.1250 _{-0.0012}^{+0.0011}$ \\
\hline
\morehorsp
Euclid-like, $5\%$ SR, T08 & $ 0.0 \pm 0.005$ & $ 0.0 \pm 0.010$ & $ 0.2002 \pm 0.0084$ & $ 0.1247 _{-0.0035}^{+0.0038}$ \\
\hline
\morehorsp
Euclid-like, $5\%$ SR, D16 & $ -0.0007 \pm 0.0047$ & $ 0.002 \pm 0.010$ & $ 0.1935 \pm 0.0088$ & $ 0.1250 \pm 0.0034$ \\
\hline
\morehorsp
Euclid-like, $10\%$ SR, T08 & $ 0.0 \pm 0.010 $ & $ 0.0 \pm 0.020 $ & $ 0.201 \pm 0.015$ & $ 0.1245 \pm 0.0051 $ \\
\hline
\hline
\morehorsp
LSST-like, $1\%$ SR, T08 & $ 0.0 \pm 0.0011 $ & $ 0.0 \pm 0.0019$ & $ 0.2001 \pm 0.0018$ & $ 0.1250 \pm 0.0011$ \\
\hline
\morehorsp
LSST-like, $5\%$ SR, T08 & $ 0.0 \pm 0.0047$ & $ 0.001 \pm 0.010$ & $ 0.1995 \pm 0.0087$ & $ 0.1251 _{-0.0046}^{+0.0044}$ \\
\hline
\morehorsp
LSST-like, $10\%$ SR, T08 & $ 0.0 \pm 0.010$ & $ 0.0 \pm 0.019$ & $ 0.201 \pm 0.015$ & $ 0.1248 \pm 0.0064$ \\
\hline
\hline
\morehorsp
WFIRST-like, $1\%$ SR, T08 & $ 0.0 \pm 0.0010 $ & $ 0.0 _{-0.0019}^{+0.0022}$ & $ 0.1999 _{-0.0020}^{+0.0022}$ & $ 0.1250 \pm 0.0013$ \\
\hline
\morehorsp
WFIRST-like, $5\%$ SR, T08 & $ 0.0 \pm 0.0049$& $ 0.0 \pm 0.010$ & $ 0.200 \pm 0.010$ & $ 0.1249 \pm 0.0053$ \\
\hline
\morehorsp
WFIRST-like, $10\%$ SR, T08 & $ 0.0 \pm 0.010$ & $ 0.0 \pm 0.019$ & $ 0.200 \pm 0.018$ & $ 0.1249 \pm 0.0077$ \\
\hline
\end{tabular}}
\caption{\footnotesize{We report the $68\%$ c.l. for the scaling relation parameters when varying the EoS for dark energy. We report results for the Euclid-like, LSST-like and WFIRST-like simulated experiments, at different scaling relation accuracies. Only for the Euclid-like experiment, we show the comparison between the T08 and D16 mass function formulations.}}
\label{tab:w0wa_sr}
\end{center}
\end{table*}

\begin{figure*}[!ht]
\centering
\includegraphics[scale=0.3]{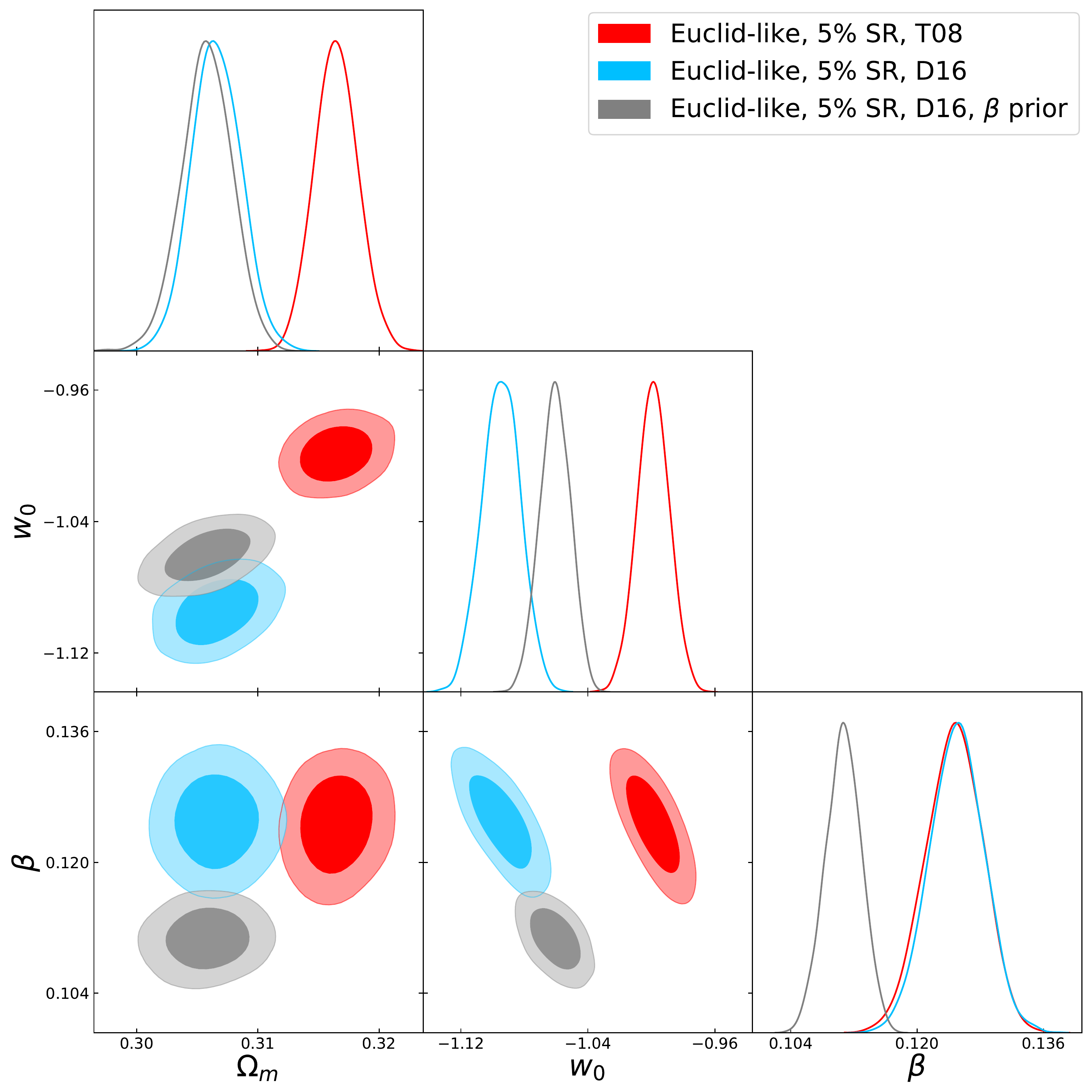}
\caption{\footnotesize{Constraints on cosmological and scaling relation parameters for the comparison between the two mass function formulations for the Euclid experiment. We show in red results when using T08 mass function and in light blue when using D16 mass function. We report also in grey the contours obtained with D16, when adopting the prior on the $\beta$ parameter from the $\Lambda$CDM analysis.}}
\label{fig:Euclid_MF_w0wa_all}
\end{figure*}

We now compare the results when adopting two different mass function formulations. As in the previous section, we show and discuss the results for the Euclid-like experiment, assuming a $5\%$ accuracy on the scaling relation parameters. 
In Fig.~\ref{fig:Euclid_MF_w0wa_all} we show a selection of cosmological and scaling relation parameters results for T08 (red contours) and D16 (light blue contours) mass function. On the one hand, we stress that in this scenario the constraints on the scaling relation parameters are consistent for the two mass function formulations. We show in Fig.~\ref{fig:Euclid_MF_w0wa_all} only the results for $\beta$, as a comparison for the shift noted in the $\Lambda$CDM scenario.
 
On the other hand, when adopting D16 mass function, we find a value of the $w_0$ parameter that is almost $8 \, \sigma$ inconsistent with the standard value $w_0 = -1$. We recall here that changing the dark energy EoS (through a shift of the $w_0$ parameter) produces changes to the redshift evolution of the growth factor and hence to the final cluster number counts. It is possible, therefore, that this shift on $w_0$ is again mimicking a different redshift evolution for the two mass functions.

This can be also seen on the results for the matter density $\Omega_m$. When adopting D16 mass function, the constraints for $\Omega_m$ are shifted towards lower values.
This shift compensates the fact that D16 mass function predicts larger cluster counts at high redshift.

This trend is also marginally visible in the results for the $\Lambda$CDM scenario (as shown in Fig.~\ref{fig:MF_LCDM_Euclid}), even though in this case this effect is mainly accounted for in the shift of the $\beta$ parameter. 

In order to further check this behaviour and the impact of the redshift evolution for the scatter, we test what happens when forcing $\beta$ to lower values. In particular, we adopt, as a prior, the constraints obtained for the $\Lambda$CDM scenario, i.e. $\beta = 0.1056  \pm 0.0025$.
We report the results in Fig.~\ref{fig:Euclid_MF_w0wa_all}, grey contours. The lower value of $\beta$ is moving the constraints on $w_0$ towards higher values, confirming the interplay of these two parameters in describing the redshift dependence on the cluster number counts.

\section{Discussion}\label{sec:discussion}

The calibration of the scaling relations, the mass function and the selection function are crucial issues when dealing with the cosmological analysis of galaxy clusters. Lots of efforts have been focused on this analysis in the last years from the international community. 

From currently available cluster catalogs, the calibration of scaling relations between cluster mass and survey observables emerges as the main source of systematic uncertainties, while the calibration of the mass function and the selection function provide subdominant impact, of the order of few percent.

The calibration of the scaling relations relies on the tight interplay between cosmology and astrophysics and is usually obtained exploiting multi-frequency observations. A proper calibration is based on the evaluation of the cluster mass and on the implementation of the relation between this mass and the survey observable

For the mass evaluation, depending on the frequency range used to detect the clusters, different mass proxies can be considered. For instance, observations in X-rays and in mm wavelengths target the hot gas in clusters and therefore make use of properties of the intra-cluster medium as mass proxies. For the observations in the optical regime, it is possible to use galaxy kinematics or weak lensing observations. These different methods may provide up to $20\%$ uncertainties on the mass evaluation. We refer the reader to the extensive discussion in \cite{Pratt:2019cnf}.

The calibration of the entire relation with the survey observable is usually obtained on a limited number of objects and is then applied to the entire cosmological sample. This approach is based on the assumption that the subsample used for the calibration is actually representative of the total cosmological sample. Furthermore, it necessitates of the understanding of how the total sample maps the underlying population, i.e. a proper description for the selection function. We refer again the reader to the full discussion in \cite{Pratt:2019cnf}.

Future surveys will provide access to cluster catalogs with $\sim 10^5$ elements. This large amount of data will nail down the impact of statistical uncertainties in the cosmological analysis, leaving the results on cosmological constraints to be fully dominated by systematic uncertainties. Therefore, apart from the mass calibration, the full characterisation of the mass function and selection function will be fundamental in order to exploit at best the cosmological constraining power of the future surveys.

In this work, we focus on the effect of improved precision on the calibration of the scaling relations and the comparison between two different evaluations of the mass function, from T08 and D16. 
Through an MCMC analysis, we forecast how the characterisation of these ingredients impact the cosmological results from future optical and near-IR galaxy surveys, comparing results for a Euclid-like, an LSST-like and a WFIRST-like experiment.

In general we find that, on the one hand, increasing the precision on the scaling relation parameters improve the constraining power. On the other hand, the evaluation of the mass function emerges as a dominant source of systematic. We perform this comparative analysis assuming a $5\%$ accuracy on the scaling relation parameters. 
We highlight that, from the comparison between T08 and D16, we see the interplay of the assumed models for the scaling relations and the mass function in the redshift evolution of the cluster number counts. 
We model the scaling relations with the mass bias and the scatter to be redshift dependent, through the $\alpha$ and $\beta$ parameters. 
When analysing the $\Lambda$CDM scenario, the comparison between the two mass functions provides consistent results on the cosmological parameters, while we obtain a $7 \, \sigma$ difference on the $\beta$ parameter. This shift encodes the different redshift evolution of T08 and D16 and in particular the fact that D16 seems to predict more clusters at higher z. 
When considering a varying EoS for dark energy, we find that this different redshift evolution is mimicked by a shift on the matter density $\Omega _m$ and the dark energy parameter $w_0$.

From the extensive discussion in D16, we recall here that the two mass function evaluations are consistent, within few percent, in the intermediate mass range, while larger differences arise when moving to more massive systems (see also the discussion in the Appendix of D16). In this case, the precision of the fit of the mass functions can be strongly affected by the resolution of the simulations used to evaluate the fitting formulas. 

Further differences may arise from the choice of the threshold used in the analysis. In our case, we consider galaxy clusters at $\Delta = 200 \rho_c$, which is shown to provide less agreement between the two formulations.
Furthermore, we note that the fitting formula used for D16 has been calibrated in the redshift range up to $z\lesssim 1.25$, while the one for T08 up to $z<2.5$. Finally, differences between the two mass functions can be due to the general calibrations that has been adopted, e.g. from the assumed cosmology in the simulations, from initial conditions, as discussed e.g. in \cite{10.1093/mnrasl/slt079}.

We stress that the impact of the choice of the mass function is different among the three experiments, due to the diverse covered mass and redshift range and distributions, as described in section \ref{sec:exp}.
In particular, for the Euclid-like and LSST-like experiments, the different evaluations provide a shift up to $1.6 \, \sigma$ along the degeneracy line in the $(\Omega_m,\sigma_8)$ plane, apart from the $\sim 7 \sigma$ shift on the $\beta$ parameter. On the contrary, when considering the WFIRST-like experiment, we recover consistent constraints for the cosmological parameters and only a $\sim 3.4 \, \sigma$ shift on the $\beta$ parameter, in the $\Lambda$CDM scenario.

We recall here that in the cosmological analysis of current cluster samples the scaling relations are calibrated to an accuracy of $\sim 10 - 20 \%$, providing uncertainties on the cosmological constraints ranging between $5\%$ and $20\%$. As mentioned above, these large errors do not allow to properly quantify the impact on the cosmological results of the mass function evaluation. 
Furthermore, we stress that the different scaling relation parameters are not known with the same accuracy. These analysis 
usually encode the redshift dependence for the scaling relations only with the self-similar scenario evolution, not adding, e.g., a redshift dependence for the mass bias or the scatter, as we are testing in our analysis.

Therefore, we highlight that the precise modelling of the cluster counts redshift evolution emerges as a fundamental step to infer cosmological constraints. Indeed, given the interplay between scaling relations and mass function, it is necessary to calibrate both on the same cluster sample, spanning a large range in mass and redshift. This is important especially to reach high accuracy for the determination of the redshift evolution of the scatter and the mass bias.

We conclude mentioning that in this analysis we did not take into account the impact of the modelling of the selection function, although it represents a fundamental part of the cluster cosmological pipeline. Nevertheless, a proper description of the cluster selection process is strictly related the the final experimental characteristics and scanning strategy. We decide therefore to model it as redshift-dependent selection in mass and focus the analysis on the interconnected impact of the scaling relations and mass function.

\section{Conclusions}\label{sec:conclusions}

We analyse the impact of the calibration for the scaling relations and the mass function on the cosmological constraints inferred from galaxy clusters detected with future optical and near-IR surveys. We perform the forecast analysis through a Monte Carlo Markov chain approach.

We model the experimental setup for three surveys, spanning different mass and redshift range and covering different areas of the sky. We focus on a Euclid-like, an LSST-like and a WFIRST like survey. In general, the Euclid-like and LSST-like experiments provide consistent results, while the WFIRST-like experiment produces wider constraints, mainly due to the substantially smaller observed sky area. 

For the scaling relations, we compare results for a $10\%$, a $5\%$ and a $1\%$ accuracy on the calibration of the parameters used to describe them. For the mass function, we compare the evaluation from \cite{Tinker:2008ff} and \cite{Despali:2015yla}.

We first analyse the impact of these modelling in the $\Lambda$CDM scenario. As expected, the increasing accuracy on the scaling relation parameters provides improved constraints on the cosmological parameters. 

The higher accuracy in the scaling relation calibration exposes the impact of the mass function evaluation, while the latter has only subdominant effects in cluster cosmological analysis from current data.
In our analysis, the effect of the two different mass function implementations is mainly seen in the results from the Euclid-like and LSST-like surveys, because of their more accurate constraints. In particular, the two implementations result in a shift up to $1.6 \, \sigma$ in the $(\Omega_m,\sigma_8)$ plane and a discrepancy of $\sim 7 \, \sigma$ in the redshift dependence for the scatter of the scaling relations. These results might be related to a different redshift evolution of the mass functions. 

This hint for a different redshift evolution is confirmed when considering a time-dependent EoS for dark energy, $w = w_0 + (1-a)w_a$. Indeed, when adopting D16 in the analysis, we find the $w_0$ parameter to be in $\sim 8 \, \sigma$ tension with the standard $-1$ value.
This implies changes in the redshift evolution of the growth factor and therefore in the final redshift distribution of cluster counts.

We conclude therefore that, a part from the well known mass calibration problem, a proper evaluation of the mass function emerges as a fundamental issue in the cluster cosmology, especially in view of future, large surveys.

\begin{acknowledgements}
      LS acknowledges support from the postdoctoral grant from Centre National d'\'Etudes Spatiales (CNES) and from the ERC-StG ‘ClustersXCosmo’ grant agreement 716762.
      The authors thank Tiago Batalha de Castro and St\'ephane Ili\'c for useful discussions.
      This project has received funding from the European Research Council (ERC) under the European Union's Horizon 2020 research and innovation programme grant agreement ERC-2015-AdG 695561.
      
\end{acknowledgements}

\bibliographystyle{aa} 
\bibliography{references} 

\begin{appendix}
\section{Results for the LCDM scenario}\label{sec:app1}

\begin{figure*}[b]
\centering
\includegraphics[scale=0.3]{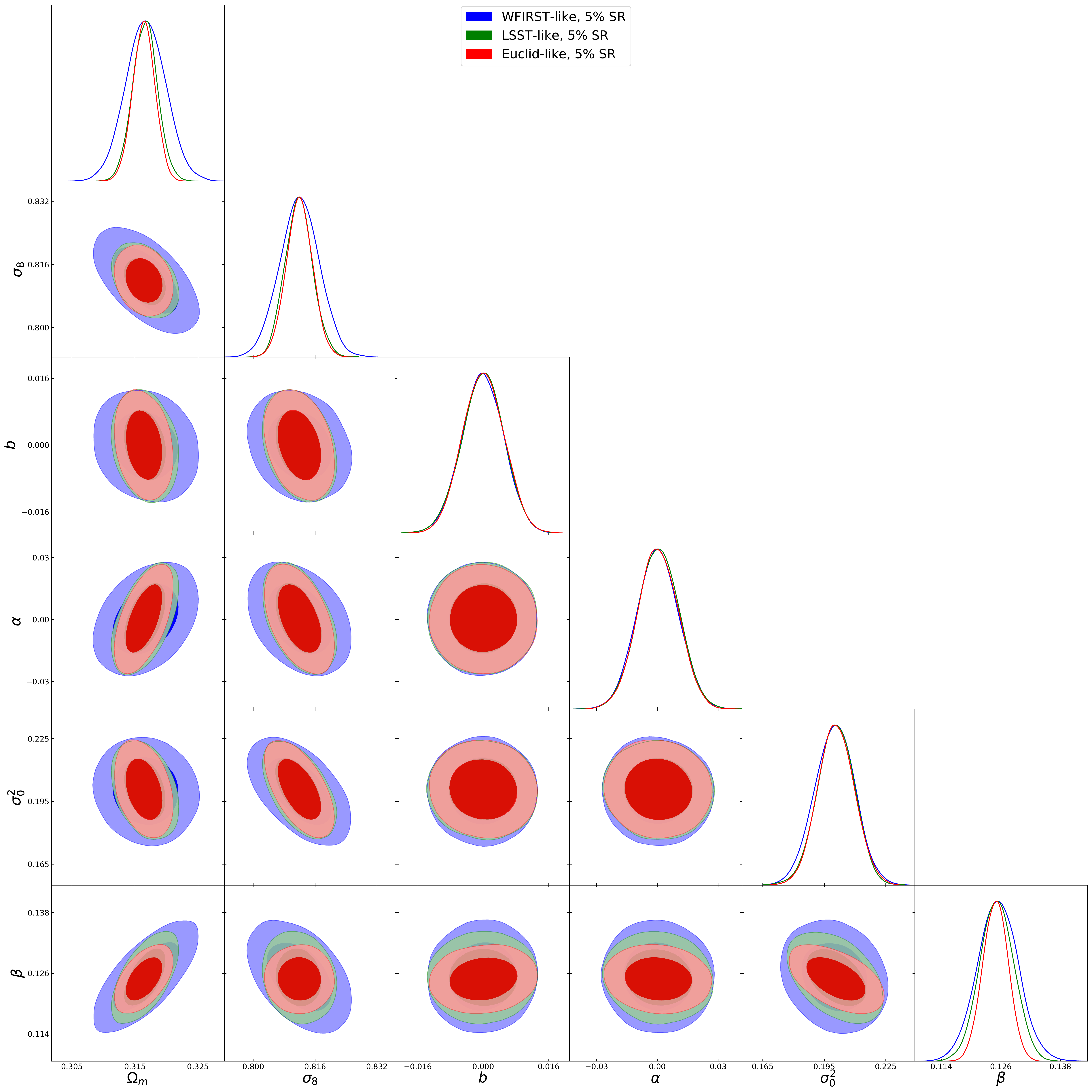}
\caption{\footnotesize{Constraints on cosmological and scaling relation parameters for the comparison between the three different experiments: WFIRST-like (blue filled contours), LSST-like (green filled contours) and Euclid-like (red filled contours). We report results when considering a $5\%$ error on the scaling relation parameters and the T08 mass function.}}
\label{fig:5SR_LCDM_allexp}
\end{figure*}

We show in the triangular plot in Fig.~\ref{fig:5SR_LCDM_allexp} the one- dimensional and two-dimensional probability distributions for
the cosmological and scaling relation parameters. We report constraints obtained with a $5\%$ accuracy on the scaling relation calibration, comparing results for the Euclid-like, LSST-like and WFIRST-like experiments.

\end{appendix}

\end{document}